# Magnetic Scattering

# Jeffrey W. Lynn[1] and Bernhard Keimer[2]

[1]NIST Center for Neutron Research

National Institute of Standards and Technology

Gaithersburg, MD 20899-6102 (USA)

[2]Max-Planck Institute for Solid State Research

Heisenbergstrasse 1

D-70569 Stuttgart (Germany)

*in*

# *Handbook of Magnetism*
edited by Michael Coey and Stuart Parkin

Contents

## I. Introduction

Scattering measurements provide essential information about the intrinsic electronic interactions in magnetic materials. Traditionally neutron scattering has been the unique magnetic scattering tool, [1] [2] but that situation has changed recently following remarkable advances in resonant x-ray scattering techniques. [3] Both techniques collect data as functions of the energy and the momentum transferred from the spin system to the neutron or photon beam. The resulting five-dimensional data sets serve as powerful probes of magnetic materials. Elastic scattering elucidates the magnetic configuration, direction of the spins, symmetry of the magnetic state, spatial distribution of the magnetization density, and dependence of the order parameter on thermodynamic fields such as temperature, pressure, magnetic and electric fields. Inelastic scattering determines the energies of the fundamental excitations, which can be used to elucidate the nature and strength of the exchange interactions.

The information provided by x-ray and neutron scattering is largely complementary. [4] For instance, resonant elastic x-ray scattering can be used to measure magnetic order parameters in an element-specific manner by tuning the x-ray energy to an electronic transition, and the enormous photon flux at modern synchrotron facilities allows measurements of very small samples and films as thin as a single unit cell. [5] [6] The magnetic dynamics of such samples can be explored with Resonant Inelastic x-ray Scattering (RIXS), [7] and pump-probe techniques open new avenues of research into non-equilibrium phenomena. Elastic neutron scattering, on the other hand, provides quantitative information about the magnitude of magnetic order parameters that is difficult to obtain with x-rays, and the energy resolution offered by inelastic magnetic neutron scattering is orders-of-magnitude finer than can be currently achieved in RIXS experiments. Both techniques can also measure the lattice structure and dynamics, as reviewed elsewhere in this volume, and those cross sections can be uniquely distinguished from magnetic scattering by polarization techniques. [8, 9]

In this chapter, we describe how to employ neutron and x-ray scattering to explore the magnetism of materials, paying particular attention to the complementarity of both techniques.

## II. Magnetic Neutron Diffraction Technique

Magnetic neutron scattering originates from the neutron's magnetic dipole moment. As a spin-½ particle, the neutron carries a magnetic dipole moment of -1.913 nuclear magnetons that interacts with the unpaired electrons in the sample, either through the dipole moment associated with an electron's spin or via the orbital motion of the electron. The strength of this magnetic interaction is comparable to the neutron-nuclear interaction. The magnetic scattering cross-section reveals the magnetic structure and dynamics of materials over wide ranges of length scale and energy. Magnetic neutron scattering plays a central role in determining and understanding the microscopic properties of a vast variety of magnetic systems – from the fundamental nature, symmetry, and dynamics of magnetically ordered materials to elucidating the magnetic characteristics essential in technological applications.

One traditional role of magnetic neutron scattering has been the measurement of magnetic Bragg intensities in the magnetically ordered regime. Such measurements can be used to determine the spatial arrangement and directions of the atomic magnetic moments, the atomic magnetization density of the individual atoms in the material, and the value of the ordered moments as a function of external parameters such as temperature, pressure, and applied magnetic or electric fields. These types of measurements can be carried out on single crystals, powders, thin films, and artificially grown multilayers, and often the information collected can be obtained by no other experimental technique. For magnetic phenomena that occur over length scales that are large compared to atomic distances, the technique of magnetic Small Angle Neutron Scattering (SANS) can be applied. This is an ideal technique to explore domain structures, long wavelength oscillatory magnetic states, vortex structures in superconductors, skyrmions, nanomagnets, and other spatial variations of the magnetization density on length scales from 1 to 1000 nm. Another specialized technique is neutron reflectometry, which can be used to investigate the magnetization profile in the near-surface regime of single crystals, as well as the magnetization density of thin films and multilayers. This particular technique has enjoyed dramatic growth during the last decade or so due to the rapid advancement of atomic-layer deposition capabilities.

The cross-section for magnetic Bragg scattering can be written as [10]

$$I_M(\mathbf{g}_{hkl}) = C \left( \frac{\gamma e^2}{2mc^2} \right)^2 M_\mathbf{g} A(\theta_B) |F_M(\mathbf{g}_{hkl})|^2 \qquad (1)$$

where $I_M$ is the integrated intensity for the magnetic Bragg reflection located at the reciprocal lattice vector $\mathbf{g}_{hkl}$, the neutron-electron coupling constant in parentheses is $-0.27 \times 10^{-12}$ cm, $C$ is an instrumental constant which includes the resolution of the measurement, $A(\theta_B)$ is an angular factor which depends on the method of measurement (sample angular rotation, $\theta:2\theta$ scan, etc.), and $M_g$ is the multiplicity of the reflection (for a powder sample). The magnetic structure factor $F_M(\mathbf{g})$ is given in the general case by [1] [11]

$$F_M(\mathbf{g}_{hkl}) = \sum_{j=1}^{N} e^{i\mathbf{g}\cdot\mathbf{r}_j} \hat{\mathbf{g}} \times \left[ \mathbf{M}_j(\mathbf{g}) \times \hat{\mathbf{g}} \right] e^{-W_j} \qquad (2)$$

where $\hat{\mathbf{g}}$ is a unit vector in the direction of the reciprocal lattice vector $\mathbf{g}_{hkl}$, $\mathbf{M}_j(\mathbf{g}_{hkl})$ is the vector form factor of the $j^{th}$ ion located at $\mathbf{r}_j$ in the unit cell, $W_j$ is the Debye-Waller factor that accounts for the thermal vibrations of the $j^{th}$ ion, and the sum is over all (magnetic) atoms in the unit cell. The triple cross product originates from the vector nature of the dipole-dipole interaction of the neutron with the electron. A quantitative calculation of $\mathbf{M}_j(\mathbf{g})$ in the general case involves evaluating matrix elements of the form $\langle \pm | 2Se^{i\mathbf{g}\cdot\mathbf{R}} + O | \pm \rangle$, where $S$ is the electron spin operator, $O$ is the symmetrized orbital operator introduced by [12], and $|\pm\rangle$ represents the angular momentum state. This can be quite a complicated angular-momentum computation involving all the electron orbitals in the unit cell, but has the simple result *that only the components of the magnetic moment that are perpendicular to $\mathbf{g}_{hkl}$ (or more generally the wave vector $\mathbf{K}$) contribute to the scattering*. Often the atomic spin density is *collinear*, by which we mean that at each point in the spatial extent of the electron's probability distribution, the atomic magnetization density points in the same direction. In this case the direction of $M_j(\mathbf{g})$ does not depend on $\mathbf{g}$, and the form factor is just a scalar function, $f(\mathbf{g})$, which is simply related to the Fourier transform of the magnetization density. The free-ion form factors have been tabulated for essentially all the magnetic elements [see, for example, https://www.ill.eu/sites/ccsl/ffacts/ffachtml.html]. Note that for x-ray scattering the form factor for charge scattering corresponds to the Fourier transform of the total charge density of all the electrons, while in the magnetic neutron case it is the transform of the "magnetic" electrons only, which are the electrons whose spins are unpaired. Recalling that a Fourier transform inverts the relative size of objects, the magnetic form factor typically decreases much more rapidly with $|\mathbf{g}_{hkl}|$ than for the case of x-ray charge scattering since the unpaired electrons are usually the outermost ones of the ion. This dependence of the scattering intensity on $f(\mathbf{g})$ is a convenient way to distinguish magnetic cross-sections from nuclear cross-sections, where the equivalent of the form factor is just a constant since the nucleus ($\approx 10^{-5}$ Å) looks like a point particle to a thermal/cold neutron (see below the nuclear coherent scattering amplitude $b$ in Eq. (6)).

If in addition to the magnetization density being collinear, the magnetic moments in the ordered state point along a unique direction (i.e. the magnetic structure is a ferromagnet, or a simple + - + - type antiferromagnet), then the square of the magnetic structure factor simplifies to

$$|F_M(\mathbf{g})|^2 = \left\langle 1 - \left(\hat{\mathbf{g}} \cdot \hat{\boldsymbol{\eta}}\right)^2 \right\rangle \left| \sum_j \eta_j \left\langle \mu_j^z \right\rangle f_j(\mathbf{g}) e^{-W_j} e^{i\mathbf{g}\cdot\mathbf{r}_j} \right|^2 \quad (3)$$

where $\hat{\boldsymbol{\eta}}$ denotes the (common) direction of the ordered moments and $\eta_j$ the sign of the moment ($\pm 1$), $\left\langle \mu_j^z \right\rangle$ is the average value of the ordered moment in thermodynamic equilibrium at ($T$, $H$, $P$, ...), and the orientation factor $\left\langle 1 - \left(\hat{\mathbf{g}} \cdot \hat{\boldsymbol{\eta}}\right)^2 \right\rangle$ represents an average over all possible domains. If the magnetic moments are the same type, then this expression further simplifies to

$$|F_M(\mathbf{g})|^2 = \left\langle 1 - \left(\hat{\mathbf{g}} \cdot \hat{\boldsymbol{\eta}}\right)^2 \right\rangle \left\langle \mu^z \right\rangle^2 f^2(\mathbf{g}) \left| \sum_j \eta_j e^{i\mathbf{g}\cdot\mathbf{r}_j} e^{-W_j} \right|^2 \quad . \quad (4)$$

We see from these expressions that neutrons can be used to determine several important quantities; the location of magnetic atoms in the unit cell and the spatial distribution of their magnetic electrons; the dependence of $<\mu^z>$ on temperature, field, pressure, or other thermodynamic variables, which is directly related to the order parameter for the phase transition (e.g. the sublattice magnetization). Often the preferred magnetic axis $\hat{\eta}$ can also be determined from the relative intensities. Finally, the scattering can be put on an absolute scale by internal comparison with the nuclear Bragg intensities $I_N$ from the same sample, given by

$$I_N(\mathbf{g}) = CM_\tau A(\theta_B) |F_N(\mathbf{g})|^2 \tag{5}$$

with

$$|F_N(\mathbf{g})|^2 = \left| \sum_j b_j e^{i\mathbf{g}\cdot\mathbf{r}_j} e^{-W_j} \right|^2 . \tag{6}$$

Here $b_j$ is the coherent nuclear scattering amplitude for the $j^{th}$ atom in the unit cell, and the sum is over all atoms in the unit cell. Typically the nuclear structure is known accurately and $F_N$ can be calculated, whereby the saturated value of the magnetic moment in Bohr magnetons can be obtained.

There are several ways that magnetic Bragg scattering can be distinguished from the nuclear scattering from the structure. Above the magnetic ordering temperature all Bragg peaks are nuclear (structural) in origin, while as the temperature drops below the ordering temperature the intensities of the magnetic Bragg peaks rapidly develop, and for unpolarized neutrons the nuclear and magnetic intensities simply add. If these new Bragg peaks occur at positions that are distinct from the nuclear reflections, then it is straightforward to distinguish magnetic from nuclear scattering. In the case of a ferromagnet, however, or for some antiferromagnets which contain two or more magnetic atoms in the chemical unit cell, these Bragg peaks can occur at the same position. One standard technique for identifying the magnetic Bragg scattering is to make one diffraction measurement in the paramagnetic state well above the ordering temperature, and another in the ordered state at the lowest temperature possible, and then subtract the two sets of data. In the paramagnetic state the (free ion) diffuse magnetic scattering is given by [5,6]

$$I_{Para} = \frac{2}{3} C \left( \frac{\gamma e^2}{2mc^2} \right)^2 p_{eff}^2 f(\mathbf{K})^2 \tag{7}$$

where $p_{eff}$ is the effective magnetic moment (= $g[J(J+1)]^{1/2}$ for a free ion). This is a magnetic incoherent cross-section, and the only angular dependence is through the magnetic form factor $f(\mathbf{K})$. Hence this scattering looks like "background". There is a sum rule on the magnetic scattering in the system, though, and in the ordered state this diffuse scattering shifts into the coherent magnetic Bragg peaks and magnetic excitations. A subtraction of the high temperature data (Eq. (7)) from the data obtained at low temperature (Eq. (1)) will then yield the magnetic Bragg peaks, on top of a deficit (negative) of scattering away from the Bragg peaks due to the disappearance of the diffuse paramagnetic scattering in the ordered state. On the other hand, all the nuclear cross-sections usually do not change significantly with temperature (apart from the Debye-Waller factor $e^{-2W}$), and hence drop out in the subtraction. A related subtraction technique is to apply a large magnetic field in the paramagnetic state, to induce a net (ferromagnetic-like) moment. The zero field (nuclear) diffraction pattern can then be subtracted from the high-field pattern to obtain the induced-moment diffraction pattern.

### a. Polarized Neutron Techniques

When the neutron beam that impinges on a sample has a well-defined polarization state, then the nuclear and magnetic scattering that originates from the sample interferes *coherently*, in contrast to being separate cross-sections like Eq. (1) and Eq. (5) where magnetic and nuclear intensities just add. Polarized neutron diffraction measurements with polarization analysis of the scattered neutrons can be used to establish unambiguously which peaks are magnetic, which are nuclear, and more generally to separate the magnetic and nuclear scattering at Bragg positions where there are both nuclear and magnetic contributions. The standard polarization analysis technique is straightforward in principle [9] [13]. Nuclear coherent Bragg scattering never causes a reversal, or spin-flip, of the neutron spin direction upon scattering. Thus the nuclear peaks will only be observed in the non-spin-flip scattering geometry. We denote this configuration as (+ +), where the incident spin of the neutron is 'up' spin and remains in the up state after scattering. Non-spin-flip scattering also occurs if the incident neutron is in the 'down' state, and remains in the down state after scattering (denoted (− −)). The magnetic cross-sections, on the other hand, depend on the relative orientation of the neutron polarization ***P*** and the reciprocal lattice vector **g**. In the configuration where ***P***⊥**g**, typically half the magnetic Bragg scattering involves a reversal of the neutron spin (denoted by (− +) or (+ −)), and half does not; the details depend on the specific Hamiltonian describing the magnetism. Thus for an isotropic Heisenberg-type model the magnetic contribution to the reflection consists of the spin-flip (− +) and non-spin-flip (+ +) intensities of equal intensity. For the case where ***P***∥**g**, all the magnetic scattering is spin-flip. Hence for a pure magnetic Bragg reflection where ($S_x$,$S_y$,$S_z$) are active, the spin-flip scattering should be twice as strong as for the ***P***⊥**g** configuration.

The arrangement of having ***P***∥**g** or ***P***⊥**g** provides an experimental simplification and hence data that are straightforward to interpret. More generally, however, ***P*** and **g** can have any relative angle. This more general technique of neutron polarimetry is more difficult to realize experimentally and can complicate the interpretation of the data, but can provide additional details about the magnetic structure that cannot be obtained otherwise. [14]

### b. Polarized Neutron Reflectometry

If neutrons are incident on a surface at (very small) grazing angles the scattering can be cast in the form of a neutron 'optical potential', analogous to photons in optical fibers. For most materials the wavelength-dependent index of refraction for neutrons (and x-rays), *n*, is slightly less then unity, so that at sufficiently small angles of incidence the scattering can be described by the one-dimensional Schrödinger equation and the neutrons undergo total external reflection—the basis for neutron guides. For a simple material with a net magnetization, interference between nuclear and magnetic scattering leads to the following expression for *n* [10] [15] :

$$n^{\pm} = \left[1 - \frac{\lambda^2}{\pi} N\left(b \pm \left(\frac{\gamma e^2}{2mc^2}\right)\langle\mu\rangle\right)\right]^{1/2} \tag{8}$$

where $N$ is the number density of the material and $\langle\mu\rangle$ is the average moment. The magnetic form factor is unity since we are scattering at very small angles. Note that $Nb$ is the nuclear scattering length density for the material, and the magnetic term is the magnetic scattering length density. The critical angle below which we have mirror reflection is given by

$$\theta_C = \arcsin\left[\frac{\lambda^2}{\pi} N\left(b \pm \left(\frac{\gamma e^2}{2mc^2}\right)\langle\mu\rangle\right)\right]^{1/2} \cong \left[\frac{\lambda^2}{\pi} N\left(b \pm \left(\frac{\gamma e^2}{2mc^2}\right)\langle\mu\rangle\right)\right]^{1/2} \tag{9}$$

where $\pm$ denotes the two polarization states of the neutron. Above the critical angle the neutrons penetrate the surface, and Fourier transforming the scattering provides a quantitative measure of the structural profile and magnetic profile of the material. For thin films and multilayers the layers, substrate, and front and back surfaces produce interference effects that provide a standard and very powerful technique for determining the properties of a wide variety of magnetic materials. [16] [17]

## III.  Resonant Magnetic X-ray Diffraction Technique

Magnetic x-ray scattering was first demonstrated off resonance, that is, with photons that were not tuned to any absorption edge of the material under study. However, the non-resonant magnetic x-ray scattering cross section is so small that this technique is not useful for magnetic structure determination. Magnetic x-ray scattering has only risen to prominence when synchrotron radiation enabled experiments with photons tuned to x-ray absorption edges, where the resonant cross section can be enhanced by several orders of magnitude. [5] [6] The enhancement is greatest when the partially occupied valence shell is reached by an electric dipole-allowed transition, that is, at the $L_{2,3}$-absorption edges of transition metals with valence $d$-electrons, and at the $M_{4,5}$-absorption edges of lanthanides or actinides with valence $f$-electrons. Magnetic x-ray scattering is then activated by the strong core-hole spin-orbit coupling in the intermediate state, prior to reemission of the photon.

From an instrumental perspective, one can group magnetic x-ray scattering experiments into three categories, depending on the photon energy $E$ required to reach the respective absorption edges, namely soft ($E < 1$ keV), intermediate ($1 \leq E \leq 5$ keV) and hard ($E > 5$ keV). Whereas soft x-ray experiments use gratings to monochromate the synchrotron radiation, intermediate and hard x-ray experiments are performed with single-crystal monochromators. Because of air absorption, soft and intermediate x-ray experiments are carried out under vacuum conditions. The soft and intermediate x-ray ranges comprise the L-edges of 3$d$ (4$d$) metals and the M-edges of 4$f$- (5$f$-) electron systems, respectively. Experiments at the dipole-active L-edges of 5$d$ metals are carried out with hard x-rays, as are experiments at the K-absorption edges of $d$-electron systems and L-absorption edges of $f$-electron systems where the resonant enhancement of the magnetic cross section is weaker.

Unlike neutron scattering, resonant magnetic x-ray scattering experiments require photons with a specific energy, so that only the direction and not the magnitude of the photon momentum is adjustable. Momentum conservation yields kinematic constraints that are particularly severe for soft x-ray experiments on the important class of 3$d$ metal compounds, where simple antiferromagnetic Bragg reflections characteristic of a doubled crystallographic unit cell cannot be reached in many cases (Fig. 1 [7]). Magnetic order with larger periodicities (and correspondingly shorter reciprocal lattice vectors) can be studied by resonant x-ray diffraction, but dynamical diffraction effects can be important (see the example below). For resonant x-ray diffraction with intermediate and hard x-rays (Fig. 1), these constraints do not apply.

In contrast to magnetic neutron scattering which is generally straightforward to interpret, a complete quantitative calculation of the magnetic x-ray scattering cross section requires numerical electronic structure calculations that describe the many-body correlations in the intermediate state. In many cases, however, one is interested in the magnetic moment orientation, which can be extracted from the dependence of the scattered intensity on the photon polarization without reference to such calculations. In spherical symmetry, the scattering tensor can be expressed in the following way: [18]

$$F_j(E) = \sigma^{(0)}(E)\, \varepsilon_i \cdot \varepsilon_o^* + \sigma^{(1)}(E)\, \varepsilon_i \times \varepsilon_o^* \cdot M_j + \sigma^{(2)}(E)\left( (\varepsilon_i \cdot M_j)(\varepsilon_o^* \cdot M_j) - \frac{1}{3}\varepsilon_i \cdot \varepsilon_o^* \right) \quad (10)$$

where $M_j$ is the magnetization vector of the ion $j$, $\varepsilon_i$ and $\varepsilon_o$ are the polarization vectors of the incoming and outgoing photons, and $\sigma^{(0)}$, $\sigma^{(1)}$, and $\sigma^{(2)}$ are proportional to the x-ray absorption (XAS), x-ray magnetic circular dichroism (XMCD), and x-ray magnetic linear dichroism (XMLD) tensors, respectively. Additional terms arise from the crystal field, but they tend to be small for collinear spin structures, as long as $M$ points along a high-symmetry direction of the crystal lattice. [18]

To separate magnetic scattering from charge scattering (first term in Eq. 10), magnetic x-ray scattering experiments can be carried out in crossed linear polarization. With the caveats mentioned above, the intensity of a magnetic Bragg reflection of a collinear antiferromagnet at the reciprocal lattice vector $g$ can then be written as

$$I = \left| \sum_j e^{ig \cdot r_j} \sigma_j^{(1)}(E)\, \varepsilon_i \times \varepsilon_o^* \cdot M_j \right|^2 \quad (11)$$

where the summation runs over the magnetic unit cell. To determine the spin structure of a given material, one commonly uses the so-called "azimuthal scan" where the momentum transfer $g$ is kept fixed, and the sample is rotated such that the orientation of $M$ varies relative to the photon polarization vectors. In this way, simple spin structures can be determined based on a single Bragg reflection.

Even for simple spin structures, however, it is important to keep in mind that the spectral functions $\sigma(E)$ are tensors with properties that may be strongly influenced by the symmetry of the crystal lattice. If the site symmetry is tetragonal, for instance, the XAS spectra for light polarized in the $xy$-plane and along the $z$-axis, $\sigma^{(0)}_{xy}$ and $\sigma^{(0)}_z$, are generally different – a phenomenon known as "natural linear dichroism". $\sigma^{(1)}$ and $\sigma^{(2)}$ are also generally anisotropic.

The deviations from spherical symmetry are particularly prominent in situations where orbital order is present. An elementary example is the $Cu^{2+}$ ion with electron configuration $3d^9$ (i.e., a single hole in the $d$-electron shell). [18] Materials based on $Cu^{2+}$ usually exhibit Jahn-Teller distortions that lift the degeneracy between $d$-orbitals of $x^2$-$y^2$ and $3z^2$-$r^2$ symmetry. The lobes of these orbitals are extended in the $xy$-plane and along the $z$-axis, respectively. For instance, the cuprate high-temperature superconductors exhibit a tetragonal structure with holes in the $x^2$-$y^2$ orbital. In this case, the electric dipole selection rules prohibit excitation of a $2p$ core electron into the valence shell with $z$-polarized light, so that $\sigma^{(0)}_z = 0$ whereas $\sigma^{(0)}_{xy} \neq 0$. The selection rule completely changes the azimuthal scans, as observed in resonant elastic scattering experiments on copper-oxide compounds. [19] This example illustrates the important influence of orbital order on azimuthal scans in magnetic x-ray scattering. Proper consideration of the crystal symmetry is especially important for experiments performed with polarized incident light, but without polarization analysis of the scattered beam, because magnetic and charge scattering may then both contribute to the detected signal.

The photon energy dependence of the scattering tensor $\sigma(E)$ contains a lot of additional information, some of which can be extracted without extensive model calculations. In particular, the large enhancement of the scattering intensity at the absorption edges of magnetic metal atoms gives rise to the element sensitivity of magnetic x-ray scattering, which is particularly useful for multinary compounds and for magnetic multilayers with different magnetic species. In principle, resonant magnetic x-ray scattering is also sensitive to the valence state of metal ions, which can be inferred from the maximum of $\sigma(E)$. Resonant scattering experiments on mixed-valent compounds have indeed been reported. [20] However, the analysis and quantitative interpretation of such experiments require careful consideration of the multiplets in the intermediate state.

In the discussion so far, we have not considered the spin-orbit coupling in the valence shell, which is generally weak for $3d$ metal compounds. In $4f$ and $5f$ electron systems, however, the spin-orbit coupling is so strong that it dominates the interatomic exchange interactions, so that models of such compounds are based on firmly locked spin and orbital angular momenta. In $4d$ and $5d$ electron systems, on the other hand, the intra-atomic spin-orbit coupling turns out to be comparable to other important energy scales including the on-site Coulomb interactions and the inter-atomic exchange coupling. Comparative magnetic x-ray diffraction experiments at the $L_2$ and $L_3$ absorption edges have recently proven to be a powerful probe of the spin-orbit composition of the ground state wave function in such materials. [19]

## IV. Dynamics
### a. Inelastic Neutron Scattering Technique

Neutrons can also scatter inelastically, to reveal the magnetic fluctuation spectrum of a material over wide ranges of energy ($\approx 10^{-8} \rightarrow 1$ eV) and over the entire Brillouin zone. Neutron scattering plays a truly unique role in that it is the only technique that can directly determine the complete magnetic excitation spectrum, whether it is in the form of the dispersion relations for spin wave excitations, wave-vector and energy dependence of critical fluctuations, crystal field excitations, magnetic excitons, or moment/valence fluctuations. In the present overview we will discuss some of these possibilities.

As an example, consider identical spins *S* localized on a simple cubic lattice, with a coupling given by $-JS_i \cdot S_j$ where *J* is the Heisenberg exchange interaction between neighbors separated by the distance *a*. The collective excitations are magnons [ref. Chapter on Spin Waves]. If we have *J>0* so that the lowest energy configuration is where the spins are parallel (a ferromagnet), then the magnon dispersion along the edge of the cube (the [100] direction) is given by

$$E(q) = 8JS[\sin^2(qa/2)] \qquad (12)$$

At each wave vector *q* a neutron can either create a magnon at (*q*, E) with a concomitant change of momentum and loss of energy of the neutron, or conversely destroy a magnon with a gain in energy. The observed change in momentum and energy for the neutron can then be used to map the magnon dispersion relation. Neutron scattering is particularly well suited for such inelastic scattering studies since neutrons typically have energies that are comparable to the energies of excitations in the solid, and therefore the neutron energy changes are large and easily measured.

Additional information about the nature of the excitations can be obtained by polarized inelastic neutron scattering techniques, which are finding increasing use. The cross section for spin wave scattering from a simple Heisenberg ferromagnet is given by [1] [13] [9]

$$\left(\frac{d^2\sigma}{d\Omega^2}\right)^{\pm} = \left(\frac{\gamma e^2}{2mc^2}\right)^2 f^2(g) \frac{k'}{k} \frac{S}{2} \frac{(2\pi)^3}{V} \sum_{\mathbf{q},\mathbf{G}} (n_\mathbf{q} + 1/2 \mp 1/2) \delta(E \mp E_\mathbf{q}) \delta(\mathbf{K} \mp \mathbf{q} - \mathbf{G})$$

$$\times \left[1 + \left(\hat{\mathbf{K}} \cdot \hat{\boldsymbol{\eta}}\right)^2 \mp 2\left(\mathbf{P} \cdot \hat{\mathbf{K}}\right) \cdot \left(\hat{\mathbf{K}} \cdot \hat{\boldsymbol{\eta}}\right)\right] \qquad (13)$$

where $n_\mathbf{q}$ is the Bose thermal population factor and $\hat{\boldsymbol{\eta}}$ is a unit vector in the direction of the spins. Generally spin wave scattering is represented by the familiar raising and lowering operators $S^{\pm} = S_x \pm iS_y$, which cause a reversal of the neutron spin when the magnon is created or destroyed. These "spin-flip" cross-sections are denoted by (+ −) and (− +). If the neutron polarization *P* is parallel to the momentum transfer $\boldsymbol{K}$, $\boldsymbol{P} \| \boldsymbol{K}$, then the spin angular momentum is conserved (as there is no orbital contribution in this case). In this experimental geometry, Eq. (13) shows us that we can only create a spin wave in the (− +) configuration, which at the same time causes the total magnetization of the sample to decrease by one unit (1 $\mu_B$ for a spin-only system). Alternatively, we can destroy a spin wave only in the (+ −) configuration, while increasing the magnetization by one unit. This gives us a unique way to unambiguously identify the spin wave scattering, and polarized beam techniques in general can be used to distinguish magnetic from nuclear scattering in a manner similar to the case of Bragg scattering.

Finally, we note that the magnetic Bragg scattering is comparable in strength to the overall magnetic inelastic scattering. However, all the Bragg scattering is located at a single point in reciprocal space, while the inelastic scattering is distributed throughout the three dimensional Brillouin zone. Hence when actually making inelastic measurements to determine the dispersion of the excitations one can only observe a small portion of the dispersion surface at any one time, and thus the observed inelastic scattering is typically two to three orders of magnitude less intense than the Bragg peaks. Consequently, these are much more time consuming measurements, and larger samples are needed to offset the reduction in intensity. Of course, a successful determination of the dispersion relations yields a complete determination of the fundamental magnetic interactions in the solid.

### b. Resonant Inelastic X-ray Scattering Technique

The mechanism underlying magnetic resonant inelastic x-ray scattering (RIXS) is analogous to the one for resonant elastic scattering discussed in Section III and depicted in Fig. 1. A photon tuned to a dipole-allowed transition promotes a core electron into the partially occupied valence shell. In the intermediate state, the core-hole spin-orbit coupling induces an electronic spin-flip, so that the re-emitted photon leaves a magnetically excited state behind. Single magnetic excitations are then observable in crossed polarization, analogous to elastic magnetic scattering (Eq. 10). [7] In this sense, the relationship between elastic and inelastic resonant x-ray scattering is analogous to the one between elastic and inelastic neutron scattering. Another useful analogy is optical Raman scattering, where single magnetic excitations at $\boldsymbol{q} = 0$ can be activated by the spin-orbit coupling in the intermediate state [21] which is, however, usually much weaker than the core-hole spin-orbit coupling in RIXS. A more common Raman scattering experiment addresses bi-magnon excitations that do not involve an electronic spin-flip. Such experiments are also possible with RIXS in parallel polarization geometry. As in optical Raman scattering, however, they only determine the Brillouin-zone averaged spectrum of magnetic excitations. The unique advantage of single-magnon RIXS is that the full magnon dispersion can be determined even for single crystals of micrometer dimensions, or for atomically thin films and heterostructures.

From an instrumental perspective, RIXS experiments on magnetic excitations are challenging because the energy of the photons required to induce the atomic dipole transition ($E = 0.4$-$1$ keV for $3d$ metal L-edges) largely exceeds the typical energy of magnons in solids. A breakthrough was achieved in 2009, when the resolving power of soft x-ray RIXS instrumentation passed the threshold of $E/\Delta E \approx 10000$. This enabled the first RIXS observation of high-energy magnons in undoped layered cuprates, which exhibit an exceptionally large bandwidth of $\approx 300$ meV. [22]

Shortly thereafter, high-energy paramagnons were also observed by RIXS in superconducting cuprates [23] [24] and in iron-based high-temperature superconductors at the Fe $L_{2,3}$ edges. [25] Kinematical constraints analogous to those in resonant elastic scattering restrict these experiments to a fraction of the Brillouin zone that does not include the magnetic ordering wave vectors of the respective parent compounds. The kinematical constraints are even more severe for RIXS experiments of bi-magnon excitations in metal oxides at the oxygen K-edge ($1s$-$2p$, 0.5 eV). [26]

Parallel advances in RIXS instrumentation for hard x-rays allowed the observation of single magnons in antiferromagnetically ordered iridium oxides with $5d$ electron systems. [27] The larger resonance energies of the $2p$-$5d$ transition, with correspondingly larger photon wave vectors, allow the detection of magnons over the entire Brillouin zone. Instrumentation for RIXS at the L-absorption edges of $4d$ metals and M-edges of actinides at intermediate photon energies ($2.5 \leq E \leq 5$ eV) has only recently been developed. [28]

In contrast to inelastic neutron scattering, the theoretical description of RIXS is still under development, and several open questions are actively debated in the literature. These include the separation of spin excitations from orbital excitations in multi-orbital systems, and from charge excitations in metallic systems. This challenge is particularly severe in the iron pnictides, which are metals with multiple Fermi surfaces originating from different Fe $d$-orbitals. A complete resolution of this problem will likely require a transition to full polarization analysis in RIXS, so that the different excitation channels can be separated completely. The first experiments using RIXS polarimeters have already been reported. [29] Another open issue is the influence of the core-hole potential in the RIXS intermediate state of the valence electron system in metallic systems, where the core-hole lifetime may be comparable to intrinsic time scales of the valence electrons.

## V.  Magnetic Diffraction Examples with Neutrons

As an example of magnetic powder diffraction, the scattering from a sample of $Na_{5/8}MnO_2$ is shown in Fig. 2 [30]. This material exhibits $Mn^{3+}$ and $Mn^{4+}$ charge stripes and vacancy ordering of the Na subsystem, which results in a rather complicated low-temperature magnetic structure that can be determined from this pattern. Of course, Rietveld refinements for the crystallographic structure can be performed from the full patterns at both high and low temperatures to determine the full crystal structure; lattice parameters, atomic positions in the unit cell, site occupancies, etc., as well as the value of the ordered moment. The inset shows the temperature dependence of the magnetic peak intensity, which we see from Eq. (4) is the square of the sublattice magnetization—the order parameter of the magnetic phase transition. Note that we can identify the magnetic scattering through its temperature dependence, as magnetic Bragg peaks vanish above the Néel temperature where long range magnetic order occurs. Note also that the magnetic intensities become weak at high scattering angles as $f(g)$ falls off with increasing scattering angle.

A more elegant way to identify magnetic scattering is to employ the neutron polarization technique, particularly if the material has a crystallographic rearrangement or distortion associated with the magnetic transition. It is more involved and time-consuming experimentally, but yields an unambiguous identification and separation of magnetic and nuclear Bragg peaks. Figure 3 shows the polarized beam results for two peaks of polycrystalline $YBa_2Fe_3O_8$. [31] The top section of the figure shows the data for the *P*⊥*g* configuration. The peak on the left has the identical intensity for both spin-flip and non-spin-flip scattering, and hence we conclude that this scattering is purely magnetic in origin. The peak on the right has strong intensity for (+ +), while the intensity for (- +) is smaller by the instrumental flipping ratio. Hence this peak is a pure nuclear reflection. The center row shows the same peaks for the *P*∥*g* configuration, while the bottom row shows the subtraction of the *P*⊥*g* spin-flip scattering from the *P*∥*g* spin-flip scattering. In this subtraction procedure instrumental background, as well as all nuclear scattering cross sections, cancel, isolating the magnetic scattering. We see that there is magnetic intensity only for the low angle position, while no intensity survives for the peak on the right, unambiguously establishing that the one peak is purely magnetic and the other purely nuclear. These data also demonstrate that all three components of the angular momentum contribute to the magnetic scattering. This simple example demonstrates how the technique works; obviously it plays a more critical role in cases where it is not clear from other means what is the origin of the peaks, such as in regimes where the magnetic and nuclear peaks overlap, or in situations where the magnetic transition is accompanied by a structural distortion where the structural peaks change significantly in intensity.

When investigating the magnetic structures of new materials, it is generally best to first carry out powder diffraction experiments to establish the basic properties of the magnetic structure, assuming of course that the ordered moment is large enough to observe the magnetic Bragg peaks. Once the basics are established, on the other hand, measurements on a single crystal can provide much higher quality and more detailed information about the magnetic properties. Figure 4 shows a map of the scattering intensity in the ($h,k$,0) scattering plane at 22 K for a single crystal of the multiferroic $Co_3TeO_6$, which orders antiferromagnetically at 26 K. [32] The crystal structure is monoclinic, and we see four satellite magnetic peaks around each (integer) structural peak, indicating that the initial magnetic structure is incommensurate in both the $h$ and $k$ (and $l$ as well, it turns out [33]) directions. With further decrease of temperature a series of additional transitions are observed, detail that would be difficult to determine with a powder. At lower temperature, separate commensurate peaks develop, then there is a lock-in transition along $k$ that includes a ferroelectric order parameter, and then finally a transition into the ground state with both commensurate magnetic order and incommensurate order along $h$, $k$, and $l$. [33] [34]

The magnetic superconductor $ErNi_2B_2C$ goes superconducting at $T_C$ = 11 K, and then develops incommensurate antiferromagnetic order below $T_M$= 6 K as shown in Fig. 5. [35] The wave vector for the ordering is ($h$,0,0) with $h \approx 0.55$, with the spin direction transverse, along (0,$y$,0). Initially the magnetic order exhibits a simple sinusoidal spin-density-wave (SDW) that is transversely polarized, as shown in the bottom of the figure. As the amplitude of the SDW increases, third, fifth, and higher-order wave vector peaks develop as the wave squares up. This is the expected behavior since for localized moments entropy mandates that a simple spin density wave cannot be the ground state magnetic structure.

For any SDW structure, only odd-order peaks will have non-zero intensity due to time-reversal symmetry, because on average the net magnetization is zero. Below 2.3 K we see that a new set of *even-order* peaks is found along the ($h$,0,0) direction of ErNi$_2$B$_2$C. One possibility is that the even-order peaks are due to a structural distortion, a charge-density wave (CDW) that follows the SDW due to a magnetoelastic interaction. Hence the even-order peaks would be structural peaks and the odd-order peaks magnetic. In the present material, however, a net magnetization develops in the superconducting state in the magnetic ground state, so that the even-order peaks could be structural, magnetic, or both. To establish the nature of these peaks unambiguously polarized neutron diffraction was used, as shown in Fig. 6. The data are measured in the ($h$,0,$l$) scattering plane, with $k$ then perpendicular to the scattering plane. For **P**‖**g** the spins are perpendicular to the scattering plane and hence perpendicular to **P** and then the magnetic scattering is all spin-flip. Note that the polarization dependence of the cross sections is quite different than the YBa$_2$Fe$_3$O$_8$ example above, emphasizing that the spin-flip and non-spin-flip magnetic cross sections depend on the details of the magnetic structure. The structural scattering is always non-spin-flip. The data show that both odd-order *and* even-order are purely magnetic in this system.

For antiferromagnets there is no net magnetization produced by the magnetic ordering. When the sublattice magnetizations are not compensated and there is a net magnetization, on the other hand, the superconductivity must respond to and try to screen this magnetization. If the internally generated field is below $H_{C1}$ then the supercurrents will exactly compensate the net magnetization and the total field will be zero. If the field exceeds $H_{C2}$ then the superconductivity will be extinguished as happens in materials such as ErRh$_4$B$_4$ and HoMo$_6$S$_8$. [36] Between these two cases, vortices are expected to be spontaneously generated, and this possibility can be investigated with SANS. Figure 7 shows SANS data from a single crystal of ErNi$_2$B$_2$C. [37] The inset presents the image on the two-dimensional SANS detector, where **K**=0 is in the center. We see the expected hexagonal pattern of scattering from the vortex lattice. Below the ferromagnetic transition additional vortices spontaneously form due to the internally generated magnetic field, which adds to the applied field. To accommodate the additional vortices they rearrange themselves with a smaller lattice parameter for the vortex lattice, which is reflected by the peak of the vortex scattering moving to larger **K**. [38]

The above examples demonstrate scattering from long range magnetic order where the magnetic diffraction consists of resolution-limited Bragg peaks. But that is not always the case, and some of the best examples occur where competing magnetic interactions lead to frustration and suppress the order or prevent it completely. Arguably the best example of a frustrated lattice occurs in the cubic rare-earth ($R$) pyrochlore ($R_2$Ti$_2$O$_7$) systems where the $R$ ions occupy corner-sharing tetrahedra. [39] For $R$ = Ho, Dy, for example, the single-ion anisotropy restricts the moments to point along diagonal [111] directions, along lines that intersect the center of each tetrahedron. The ground state turns out to be with two of the moments pointing into each tetrahedron and two pointing out. But you don't know which two are in and which two are out, exactly like the hydrogen bonding in ice where two $H$ move into the oxygen in the center of the tetrahedron and bond and two move out, resulting in a macroscopic degeneracy that violates the third law of thermodynamics. The first measurement of the ground state correlations was carried out for $Ho_2Ti_2O_7$, where the observed scattering from the correlated moments agreed quite well with simulations. [40] An interesting simplification occurs for a field applied along the [111] direction, which isolates the layers and forms two-dimensional 'kagomé spin-ice'. The scattering for this case is shown in Fig. 8 for $Dy_2Ti_2O_7$, which shows the broad distributions of diffuse magnetic scattering that are in excellent agreement with Monte Carlo simulations. [41]

The ground state properties are not the only remarkable property of spin-ice, as the magnetic excitations are equally fascinating. Theory showed that these excitations, which simply consist of flipping one of the spins in a tetrahedron so that you have three pointing out and one pointing in (and in the adjacent tetrahedron three point in and one out), correspond to the creation of a magnetic monopole and anti-monopole. [42] The subsequent motion of these particles is governed by the Coulomb Hamiltonian for magnetic charges, and this scenario was subsequently confirmed by neutron scattering measurements. [43] [41] [44]

Advances in thin film deposition methods have facilitated the synthesis of complex heterostructures with atomic layer accuracy, which has enabled investigators to control the magnetic properties by tailoring the exchange interactions within and between layers. These capabilities combined with advances in experimental reflectometry techniques have made neutron scattering an essential tool to elucidate the atomic depth profile and magnetization density of thin films and multilayers. An interesting example is the multilayer oxide heterostructure consisting of the (approximately cubic) antiferromagnets $LaMnO_3$ and $SrMnO_3$, grown on a $SrTiO_3$ substrate. The structural indices of refraction for these two materials are almost identical, rendering the structural scattering practically invisible. Occasionally an extra layer of $LaMnO_3$ was deposited to dope the interface, which produced an effective composition of $La_{0.44}Sr_{0.56}MnO_3$, which is in the ferromagnetic regime. Figure 9 shows the non-spin-flip polarized neutron reflectivity data in the two polarization states, $R^{++}$ and $R^{--}$, that are sensitive to the ferromagnetism. The resulting magnetic depth profile reveals that the magnetic modulation is quite large, varying from 0.7 $\mu_B$ to 2.2 $\mu_B$, and that its period corresponds precisely to the LMO superlattice structure. [45] High angle diffraction data on the epitaxial multilayer confirmed the canted modulated spin structure of the superlattice.

## VI. Magnetic Diffraction Examples with X-rays

As an example of resonant magnetic x-ray scattering, we first highlight experiments on the antiferromagnet $Sr_2IrO_4$ with hard x-rays tuned to the Ir $L_{2,3}$ edges [46]. The crystal structure of $Sr_2IrO_4$ is composed of $IrO_2$ square lattices, closely similar to $La_2CuO_4$, the parent compound of a prominent family of high-temperature superconductors. Prior to the x-ray experiments, magnetic susceptibility measurements had suggested antiferromagnetic order with a Néel temperature of 240 K, but neutron diffraction experiments had proven difficult because of the large neutron absorption cross section of Ir, and because large single crystals could not be grown. The hard x-ray data on a crystal of sub-millimeter dimensions show multiple magnetic Bragg reflections that can be analyzed by refining the Bragg intensities according to Eq. (11) in a manner entirely analogous to magnetic neutron diffraction. The analysis revealed a canted antiferromagnetic structure in the $IrO_2$ planes, with alternating stacking in the direction perpendicular to the planes.

The photon energy dependence of the resonant magnetic x-ray scattering cross section yields additional information about the magnetic ground state of $Sr_2IrO_4$ that would be difficult to obtain with neutron diffraction, even under ideal conditions. The Ir valence electrons occupy $5d$ orbitals of *xy*, *xz*, and *yz* symmetry. For materials with $3d$ valence electrons, the crystal field lifts the degeneracy between these orbitals and quenches the orbital magnetization. In the $5d$ electron shell, however, the strong intra-atomic spin-orbit coupling can generate complex admixtures of these orbitals in the ground-state wave function, which correspond to a nonzero orbital magnetic moment. This, in turn, affects the matrix elements for the photon-induced transitions from the spin-orbit split $2p$ shell into the $5d$ shell such that the diffraction intensities at the $L_2$ and $L_3$ edges ($2p_{1/2}$-$5d$ and $2p_{3/2}$-$5d$, respectively) can become different. The strong disparity of the diffraction intensities observed experimentally (Fig. 10) [46] indicates that the orbital magnetization is largely unquenched, and that the spin and orbital components of the magnetic order parameter in $Sr_2IrO_4$ are of comparable magnitude. Similar observations have been made for other iridates. Models of magnetism in the iridates are therefore commonly expressed in terms of the total angular momentum, $J_{eff}=S+L$. For $Sr_2IrO_4$, $J_{eff} = ½$ in the ground state.

The large resonant scattering cross section, combined with the high photon flux at synchrotron beamlines and the focusing capability of advanced x-ray instrumentation, allow magnetic x-ray scattering experiments with beam dimensions well below typical magnetic domain sizes. Figure 11 provides an example of such an experiment on the layered antiferromagnet $La_{0.96}Sr_{2.04}Mn_2O_7$, which comprises alternately stacked sheets of ferromagnetically aligned Mn spins [47]. The (001) magnetic Bragg reflection of this spin array can be reached with photons tuned to the Mn $L_3$-edge. The data shown in Fig. 11 were taken with a beam of 300 nm diameter. They reveal that the diffracted intensity varies on a characteristic length scale of several microns. A detailed analysis shows that the intensity variation results from domains with different spin directions, which diffract photons with different scattering amplitude due to the photon polarization dependence of the scattering cross section (Eq. (11)). In another study, domains with different helicities in a spiral magnet were imaged by resonant diffraction with circularly polarized x-rays. The spatial resolution and imaging capabilities of magnetic x-ray scattering methods are expected to develop rapidly with the advent of coherent x-ray beams at fourth-generation synchrotron sources.

In analogy to neutron reflectometry, polarized magnetic x-ray reflectometry has recently developed into a powerful, element-sensitive probe of complex oxide thin films, heterostructures, and superlattices. As an example, we discuss resonant x-ray diffraction data on $R$NiO$_3$-based films and superlattices (where $R$ denotes a lanthanide atom). $R$NiO$_3$ perovskites exhibit a Mott metal-insulator transition as a function of the radius of the $R$ cation, which modulates the Ni-O-Ni bond angle. Recent work has shown that the metal-insulator transition can also be controlled by epitaxial strain and by spatial confinement of the conduction electron system. Antiferromagnetism with ordering vector $g$ = (¼, ¼, ¼) develops in the Mott-insulating phase. Fig. 12(top) shows azimuthal scans at the corresponding magnetic Bragg reflection taken with photons tuned to the Ni $L_3$-edge [48]. The data analysis demonstrates that the magnetic order is non-collinear, with Ni spins forming a spiral propagating along the (111) direction of the perovskite unit cell. The polarization plane of the spiral can be controlled by epitaxial strain.

Fig. 12(bottom) shows a contour map of the resonant scattering intensity from a $LaNiO_3$-$LaAlO_3$ superlattice as a function of the azimuthal angle and the momentum transfer perpendicular to the superlattice plane. [49] Strong modifications of the azimuthal-angle dependence of the intensity occur particularly under grazing-incidence or grazing-exit conditions, where the incident or scattered beams are strongly refracted at the external and internal interfaces of the superlattice. These data illustrate the possibly important influence of dynamical effects in resonant soft x-ray diffraction from thin-film structures, which go beyond the kinematic approximation usually employed in the analysis of such data.

Very recently, x-ray free-electron lasers have enabled time-resolved resonant magnetic diffraction experiments capable of imaging the real-time dynamics of magnetic order under non-equilibrium conditions. As an illustration of this emerging capability, Fig. 13 shows the time evolution of the $g$ = (¼, ¼, ¼) antiferromagnetic Bragg peak of a $NdNiO_3$ film following a THz pump pulse exciting an infrared-active phonon mode of the $LaAlO_3$ substrate [50]. As the phonon propagates from the substrate through the film, it obliterates the antiferromagnetic order in its wake on a picosecond time scale. The mechanism underlying this "non-thermal melting" phenomenon may involve transient distortions of the $NiO_6$ octahedra, which weaken the magnetic exchange interactions between Ni spins.

## VII. Spin Dynamics with Neutrons

There are many types of magnetic excitations and fluctuations that can be measured with neutron scattering techniques, such as magnons, spinons, critical fluctuations, crystal field excitations, magnetic excitons, and moment/valence fluctuations. We start with classic magnons in an isotropic ferromagnet, where the excitations are gapless and the dispersion relation is given by Eq. (12). Figure 14(left) shows a measurement for $La_{0.67}Ca_{0.33}MnO_3$, which is a colossal magnetoresistive (CMR) material. [51] The data reveal two magnon peaks at a given wave vector, one in energy gain where the neutron destroys a magnon and gains energy, and one in energy loss where a magnon is created. This is a small $q$ (long wavelength) excitation, and in fact this sample is polycrystalline rather than single crystal, and the data were collected around the (0,0,0) reciprocal lattice position. Such measurements are restricted in wave vector and energy, and are only viable for isotropic ferromagnets; otherwise the excitations fall outside the accessible experimental window dictated by momentum and energy conservation. If there is a question of whether these excitations are magnons or phonons, the polarized beam technique can be employed as shown in Fig. 14(right) for the prototypical isotropic ferromagnet amorphous $Fe_{86}B_{14}$. [52]. These data were taken with the neutron polarization $P$ parallel to the momentum transfer $K$ ($P \parallel K$). In this configuration magnons require the neutron spin direction to reverse (spin-flip), while phonons can only be observed in the non-spin-flip configuration.  For magnons we should be able to create a spin wave only in the (− +) configuration where the incident neutron moment is antiparallel to the magnetization; the scattered neutron moment is then parallel to the magnetization direction, and the magnetization is decreased by one unit by the creation of the magnon.  On the energy gain side the process is reversed and we destroy a magnon only in the (+ −) configuration.  This is precisely what we see in the data; for the (− +) configuration the spin waves can only be observed for neutron energy loss scattering (E > 0), while for the (+ −) configuration spin waves can only be observed in neutron energy gain (E < 0).  This behavior of the scattering uniquely identifies these excitations as magnons.

Expanding the sine in Eq. (12) we see that the small-$q$ dispersion relation can be written as $E_{sw} = D(T)q^2$, where $D$ is the spin wave "stiffness" constant.  The general form of the spin wave dispersion relation is the same for all isotropic ferromagnets, a requirement of the (assumed) perfect rotational symmetry of the magnetic system, while the numerical value of $D$ depends on the details of the magnetic interactions and the nature of the magnetism.  The small-$q$ dispersion relation can be readily measured, as shown in Fig. 15(left) for a single crystal of $La_{0.85}Sr_{0.15}MnO_3$, and $D(T)$ obtained. [53] The effect of temperature is to soften the average exchange interaction as the magnetization decreases, and hence the magnons renormalize to lower energies with increasing temperature as also shown Fig. 15.  With single crystals the dispersion curves can be determined in different directions and throughout the Brillouin zone, as shown in Fig. 15(right) for a number of perovskite CMR systems. [54] Such measurements enable to determine in detail all exchange interactions, rather than just the long wavelength (average) behavior.  Any gap(s) in the excitation spectrum can also be directly measured.

In addition to the magnon energies, the lifetimes of the excitations can also be determined by extracting the intrinsic widths of the excitations, both in the ground state for itinerant electron systems, and as a function of temperature.  An example of the linewidths in the ground state are shown for $La_{0.85}Sr_{0.15}MnO_3$ in Fig. 16. [53] In the simplest localized-spin model negligible intrinsic spin wave linewidths would be expected at low temperatures, while we see here that the observed linewidths are substantial at all measured wave vectors and highly anisotropic, indicating that an itinerant electron type of model is a more appropriate description for this system.  In particular, the linewidths become very large at large wave vectors.  These substantial linewidths are easy to measure with conventional instrumentation.  Insulating magnets, on the other hand, generally have much smaller linewidths and require much higher instrumental resolution to measure. Figure 16(right) shows the measured linewidths for the prototype insulating antiferromagnet $Rb_2MnF_4$. [55] Here the spin-echo triple-axis technique has been employed, which has extraordinarily good (μeV) resolution.  The theoretically calculated linewidths from spin-wave theory are shown by the solid curves at a series of temperatures, and are in quantitative agreement with the data.

One area where neutron scattering has played an essential role is elucidating the spin dynamics of the high temperature superconductors, first for the copper oxide systems [56] and more recently for the iron-based superconductors. [57] The magnetic excitations in these classes of materials extend to quite high energies—as high as ≈0.5 eV—making the measurements particularly challenging since the magnetic form factor requires that the magnitude of $K$ must be kept small, necessitating quite high incident energy neutrons.  These requirements are well matched to the time-of-flight capabilities of spallation neutron facilities where high energy neutrons are plentiful.

To illustrate the basic technique, consider the excitations from BaFe$_2$As$_2$, which is one of the antiferromagnetic 'parent' materials of the iron-based superconductors. The antiferromagnetic ordering temperature $T_N$ = 138 K, which corresponds to a thermal energy of just ≈12 meV (1 meV → 11.605 K). Yet we see from Fig. 17 that the magnons extend up to 200 meV, an order-of-magnitude higher energies than the ordering temperature represents, indicating that the system has a substantial component of low-dimensional character. [58] The in-plane dispersion relations are also quite anisotropic, even though the orthorhombic distortion away from tetragonal symmetry (that accompanies the magnetic order) is small. Another very interesting aspect of the magnetic excitations is that they have quite large linewidths at high energies, indicating that the magnetic electrons are itinerant in nature. Indeed, the iron *d*-bands where the magnetism originates cross the Fermi energy—the definition of itinerancy.

Our final neutron example concerns the spin dynamics of one-dimensional (1D) magnets, which (together with 2D magnets) have played a special role in developing a fundamental understanding of quantum magnetic systems. This is because they are theoretically more tractable and therefore enable a deeper comparison with experiment. They also entail the emergence of new types of cooperative states and their associated excitations. Arguably the most interesting case is for the spin one-half antiferromagnet chain, where quantum effects are maximal, represented by materials such as KCuF$_3$ [59] and CuSO$_4$·5D$_2$O [60] which have enjoyed a long and interesting history of investigations. The ground state turns out to be an entangled macroscopic singlet, but where the two-spin correlation function decays only algebraically, rendering long lengths of the chain to be correlated antiferromagnetically. The fundamental excitations of such a 1D system are spinons in these (isolated) spin chains, which can be considered to a first approximation as moving domain walls. Measurements of the dynamic structure factor for CuSO$_4$·5D$_2$O are shown in Fig. 18. [60] Spinons carry fractional spin, and hence these fractionalized excitations can only be created in pairs in the scattering process. Thus the lower energy part of the spectrum corresponds to two-spinon excitations and has the appearance of a simple antiferromagnetic spin wave dispersion relation. However, only 71 % of the spectral weight is contained in this two-spinon component, with essentially all the remainder being accounted for by the four-spinon contribution. Precise calculations of the dynamic structure factor for two-spinon and four-spinon scattering are also shown in Fig. 18, which account for essentially the entire measured spectral weight, and are in excellent agreement with the measurements. [60]

## VIII. Spin Dynamics with RIXS

The set of materials investigated by high-resolution RIXS is thus far limited to magnets with characteristic exchange interactions of the order of 100 meV. A milestone was set by early experiments on La$_2$CuO$_4$, the antiferromagnetic, Mott-insulating end member of a family of high-temperature superconductors, which exhibits an exceptionally large magnon bandwidth of ≈300 meV. A RIXS spectrometer with energy resolution of $\Delta E \approx$ 100 meV proved to be capable of separating these excitations from the elastic line over a substantial fraction of the Brillouin zone (Fig. 19). [23, 22] Comparison with prior inelastic neutron scattering data on the same materials demonstrated that the RIXS excitation features indeed originate from single antiferromagnetic magnons.

RIXS experiments have also revealed the persistence of high-energy paramagnon excitations in highly doped, superconducting cuprates. Based on the polarization dependence of the scattering cross section at specific scattering geometries, they can be separated from charge excitations, as shown in Fig. 20 for $YBa_2Cu_3O_{6+x}$. [29] The measurements are complementary to inelastic neutron scattering experiments, which have much higher energy resolution and can therefore access spin excitations with energies from 1-100 meV, comparable to the superconducting energy gap. The RIXS measurements, on the other hand, are more sensitive to high-energy excitations, which can also be investigated with high energy neutrons from spallation sources. The photon energy dependence of the RIXS intensity yields additional insight into the nature of these excitations. Whereas the spin excitation energy is independent of photon energy, as expected for collective modes, the spectral weight of the charge excitations shifts upon detuning the photon energy away from the L-edge resonance, signaling a broad excitation continuum. This supports models that treat collective spin excitations as mediators of unconventional superconductivity.

Hard x-ray RIXS experiments on the layered iridates have revealed magnon dispersions remarkably similar to those of the cuprates – a finding that has fueled predictions of unconventional superconductivity in the iridates. In addition to the usual low-energy magnon branches emanating from the antiferromagnetic Bragg reflections, these experiments have also revealed weakly dispersive "spin-orbit exciton" modes corresponding to spin excitations from the $J_{eff} = ½$ ground state into the $J_{eff} = 3/2$ excited state (Fig. 21). [27] Since the dispersion of these modes is controlled by the combination of the intra-atomic spin-orbit coupling, the crystalline electric field, and the inter-atomic exchange interactions, RIXS experiments are an incisive probe of the low-energy electronic structure of these materials.

Finally, to illustrate the diversity of inelastic x-ray scattering methods applied to magnetism, we highlight results of an x-ray emission spectroscopy study of iron arsenide superconductors of composition $Ca_{1-x}R_xFe_2As_2$ (where $R$ = rare earth). [61] The goal of this experiment was to elucidate the origin of a pressure-induced structural phase transition from an antiferromagnetic to a nonmagnetic state that is associated with a large volume reduction. [62] [63] To measure the local magnetic moment of the Fe ions independent of any interatomic correlations, x-ray photons were tuned to the Fe K-absorption edge (1$s$-3$d$), and the spectrum of emitted x-rays was monitored around the dipole-active $K_\beta$ emission line (2$p$-1$s$). A local moment on the Fe site induces a splitting of this line (inset of Fig. 22) whose size depends on the moment amplitude. These experiments led to the discovery of a pressure induced spin-state transition from a high-spin to a low-spin configuration of the Fe atoms. The lower volume of the low-spin Fe atoms explains the volume collapse in the nonmagnetic phase at high pressures.

## IX. Facilities and Online Information

A list of current neutron scattering facilities around the world can be found at (http://en.wikipedia.org/wiki/Neutron_research_facility). Numerical values of the free-ion magnetic form factors for neutrons can be obtained at https://www.ill.eu/sites/ccsl/ffacts/ffachtml.html. Values of the coherent nuclear scattering amplitudes and other nuclear cross-sections can be found at http://www.ncnr.nist.gov/resources/n-lengths/.

A list of current x-ray scattering facilities can be found at (http://en.wikipedia.org/wiki/List_of_synchrotron_radiation_facilities).

Values for characteristic x-ray energies and a guide to the literature on x-ray form factors can be found at http://xdb.lbl.gov/.

Energy Units: Traditionally magnetic excitations are quoted in units of meV but sometimes authors use THz, particularly for phonons in older literature. Raman and IR experimenters often use cm$^{-1}$. 1 meV → 0.24180 THz → 8.0655 cm$^{-1}$ → 11.605 K.

For a wavelength $\lambda$ = 1.54 Å the photon energy is 8.05 keV, the electron energy is 63.4 eV, and for a neutron the energy is 34.5 meV.

## X.  Summary and Future Directions

In this review we have discussed the basic characteristics of magnetic neutron and x-ray scattering and provided a number of experimental examples of how these techniques can be employed. Neutron scattering is a rather mature technique which has the advantage of being a weakly interacting probe that does not affect the properties of the sample. The source of neutrons has traditionally been steady state reactor based facilities, but this has now been complemented by the newer, pulsed spallation neutron source facilities which can offer higher peak flux than steady-state reactors. Both types of sources have many different types of spectrometers that enable magnetic investigations over many orders-of-magnitude in both spatial and time domains. In addition to new sources and new types of sources, many of the advancements in neutron techniques over the years have come from developments in how to tailor and manipulate neutrons, vast arrays of detectors, and the software to analyze and visualize the data, and this progress continues unabated. New sources and new instrumentation currently are being planned and developed, with the anticipation that measurement capabilities will be greatly increased together with an increased quantity and scale of data acquired.

Resonant x-ray scattering is a much newer technique, with high brightness that allows measurements of small bulk samples, thin films, and multilayers. It also has the advantage of being element specific as the resonance is tuned to an absorption edge. Tremendous progress in measurement capabilities has been realized in the last few years, both with magnetic diffraction and magnetic inelastic scattering. In contrast to neutron scattering which is on a solid theoretical foundation, the theoretical understanding and interpretation of magnetic x-ray scattering is undergoing considerable development, which should lead to improved interpretation of experimental data and exciting new capabilities.

The future of both techniques is brilliant, pun intended.

**Acknowledgments**

We thank our many colleagues who have collaborated with us on our own projects, where we have taken a number of the examples for our own convenience and familiarity with the systems.**Acknowledgments**

We thank our many colleagues who have collaborated with us on our own projects, where we have taken a number of the examples for our own convenience and familiarity with the systems.

# Figures

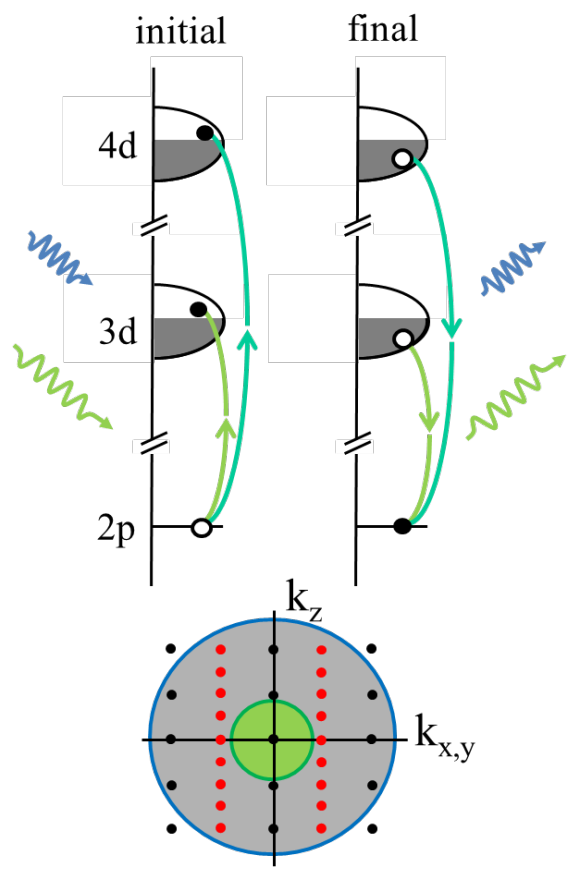

**Figure** 1. (top) Energy/density-of-states diagram illustrating RIXS with photons near the L-absorption edges of 3d (green) and 4d (blue) metals. (bottom) Reciprocal lattice of orthorhombic perovskite antiferromagnets, with structural (black) and magnetic (red) Bragg reflections. Circles indicate the maximal coverage of RIXS with photons at the Cu (green) and Ru (blue) $L_{2,3}$-edges. (top panel adapted from [7], © American Physical Society 2011).

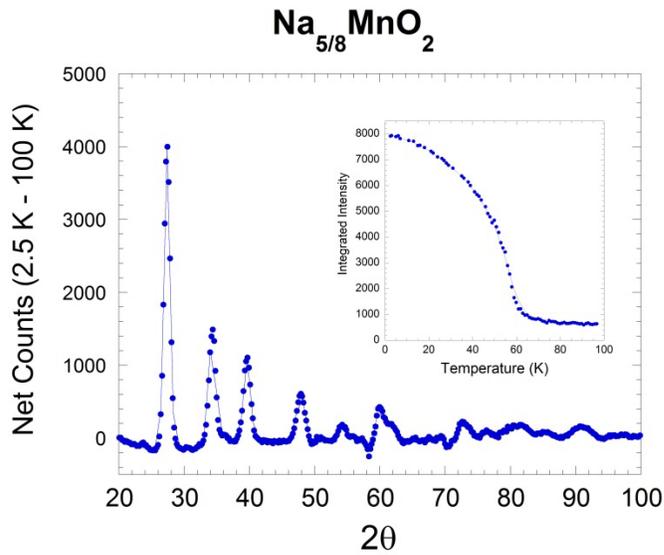

Figure 2. Magnetic diffraction pattern for Na$_{5/8}$MnO$_2$, obtained by subtracting the crystallographic diffraction pattern obtained at 100 K, above the antiferromagnetic phase transition, from the data at 2.5 K in the magnetic ground state. The structural scattering cancels in the subtraction if there is no significant change when the sample magnetically orders. The inset shows the temperature dependence of the intensity of the strongest magnetic peak, and reveals a transition temperature of ≈60 K (adapted from [30], ©Spinger Nature 2014).

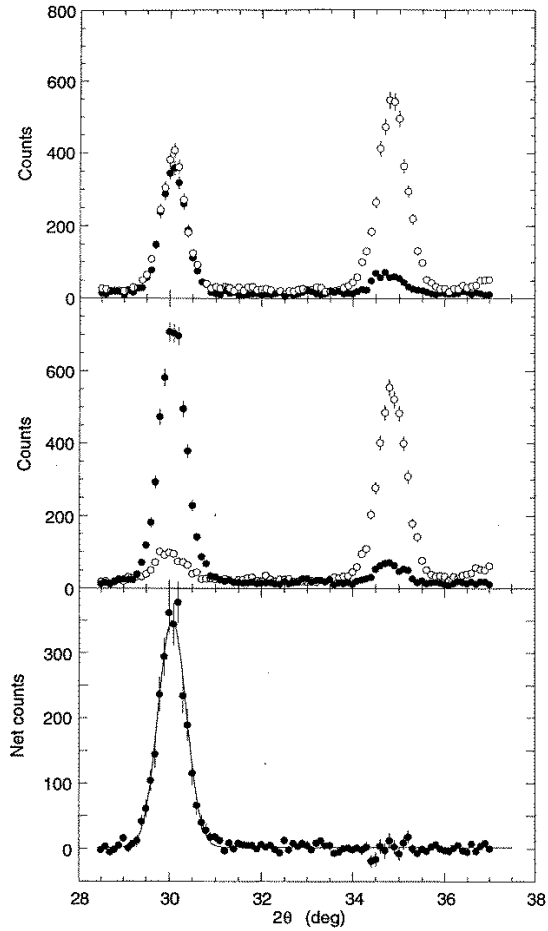

Figure 3. Polarized neutron diffraction on polycrystalline YBa$_2$Fe$_3$O$_8$. The top portion of the figure is for **P**$\perp$**g**, where the open circles show the non-spin-flip scattering and the filled circles are in the spin-flip configuration. The low angle peak has equal intensity for both cross sections, and thus is identified as a pure magnetic reflection, while the ratio of the (+ +) to (- +) scattering for the high angle peak is just the instrumental flipping ratio. Hence this is a pure nuclear reflection. The center portion of the figure is for **P**||**g**, and the bottom portion is the subtraction of the spin-flip data for the **P**$\perp$**g** configuration from the spin-flip data for **P**||**g**. Note that in the subtraction procedure all background and nuclear cross sections cancel, thereby isolating the magnetic scattering. (reprinted by permission from [31], © American Physical Society 1992).

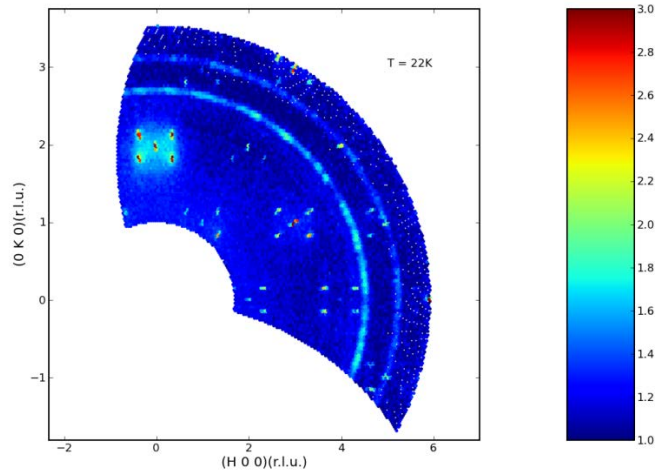

Figure 4. Neutron diffraction intensity map observed in the ($h$, $k$, 0) scattering plane of a single crystal of the multiferroic $Co_3TeO_6$. The temperature is 22 K, just below the antiferromagnetic phase transition at $T_N$=26 K. The nuclear Bragg peaks at integer positions are accompanied by four satellite magnetic reflections, indicating the development of incommensurate (ICM) magnetic order. Note that the ordering wave vector is incommensurate in both $h$ and $k$. No energy analyzer was used for these measurements so that the data are energy-integrated, and there is clear diffuse scattering surrounding the ICM peaks at this temperature originating from inelastic magnetic excitations (adapted from [32], © American Physical Society 2012).

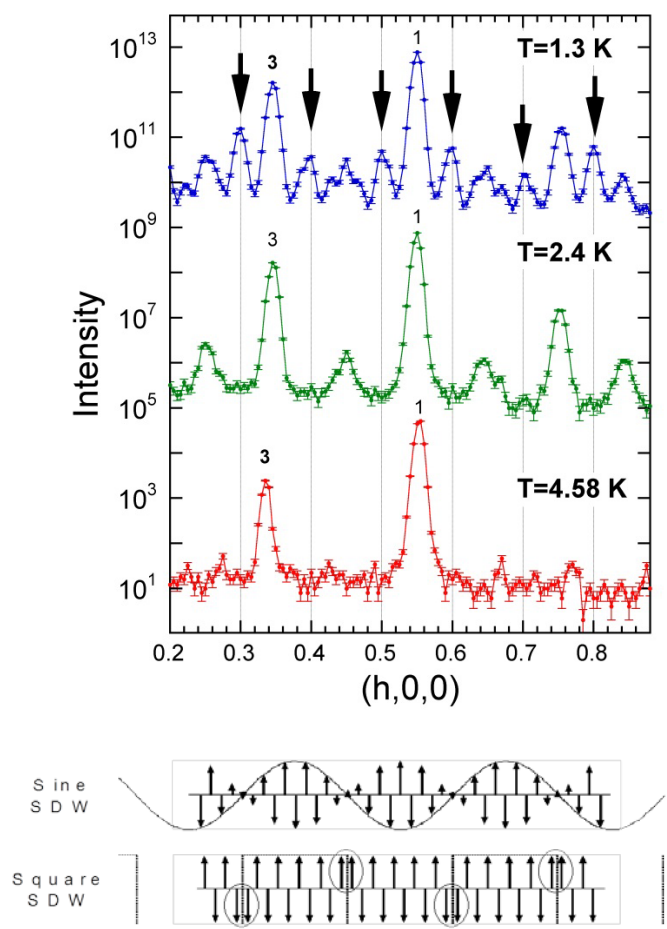

Figure 5. (top) Unpolarized neutron diffraction measurements along the ($h$,0,0) direction at 1.3 K, 2.4 K, and 4.58 K of a single crystal of ErNi$_2$B$_2$C. At 10 K no peaks are observed in this wave vector range. The data have been offset along the intensity axis for clarity. Above the weak ferromagnetic transition at 2.3 K the fundamental incommensurate peak is observed at $h$=0.55, along with higher odd-order harmonics. Below the ferromagnetic transition a new set of even-order harmonics develops, indicated by the arrows. (bottom) Schematic of the initial transversely polarized spin-density-wave, and ground state square-wave (adapted from [35], © American Physical Society 2001).

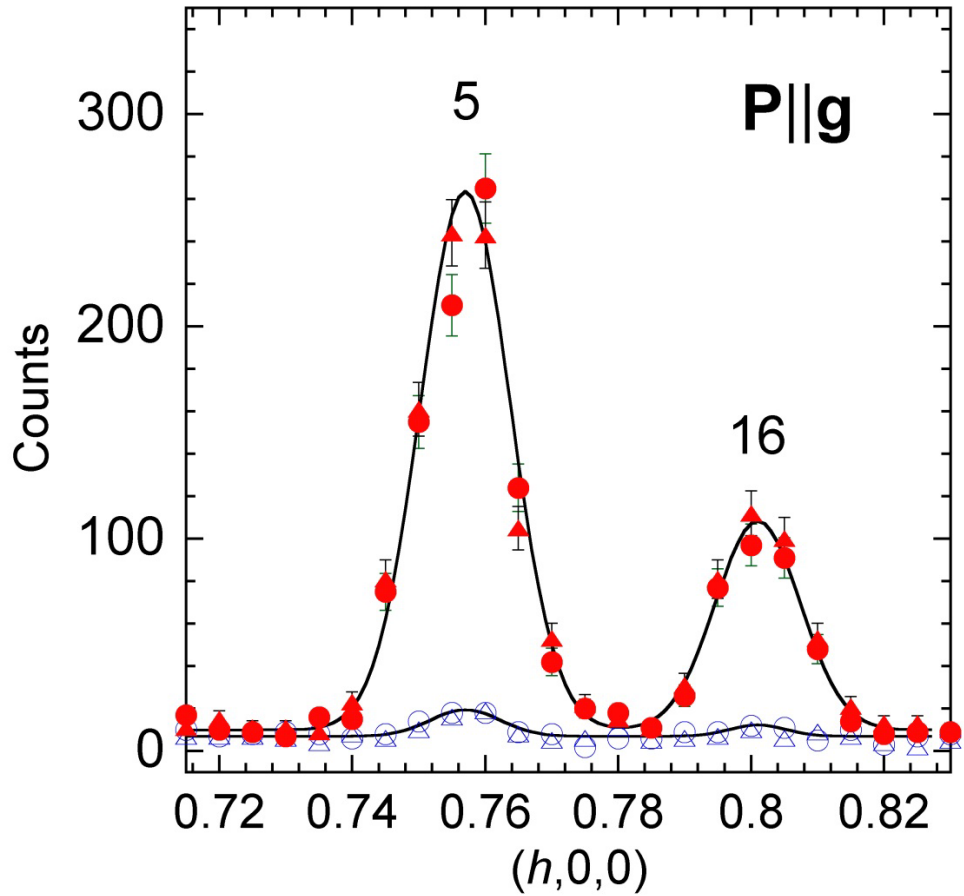

Fig. 6. Polarized neutron diffraction measurements on a single crystal of ErNi$_2$B$_2$C showing both the odd order (5$^{th}$) and even-order (16$^{th}$) harmonics for the **P**||**g** configuration. The solid circles (- , +) and solid triangles (+ , -) are spin-flip scattering, while the open circles (+ , +) and open triangles (- , -) are non-spin-flip scattering. The data demonstrate that both types of reflections are magnetic in origin, with the moment direction along the *b* axis (adapted from [35], © American Physical Society 2001).

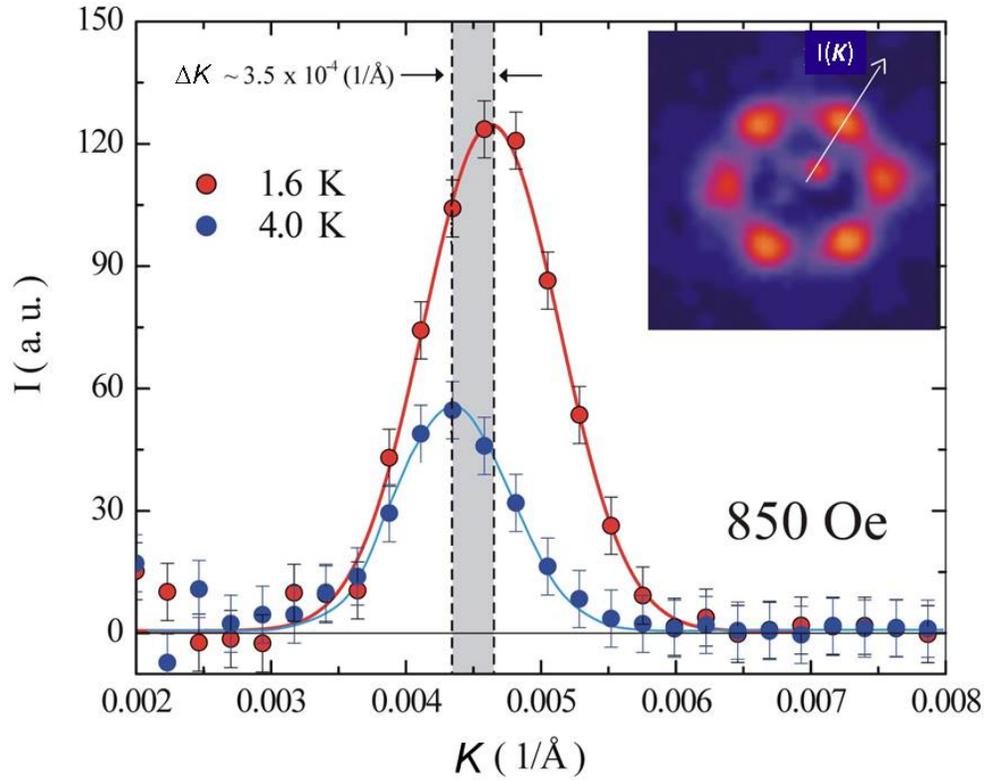

Figure 7. Radially averaged small angle neutron scattering intensity of the vortex scattering in ErNi$_2$B$_2$C vs. wave vector ***K*** at 85 mT, above and below the weak ferromagnetic transition. The shift in the peak position demonstrates that additional vortices spontaneously form as the macroscopic magnetization develops at low temperatures. The temperature dependence shows that this spontaneous vortex formation is directly related to the weak ferromagnetic transition. The inset shows vortex Bragg peaks on the two-dimensional SANS detector; **K** = 0 is in the center. (adapted from [37]).

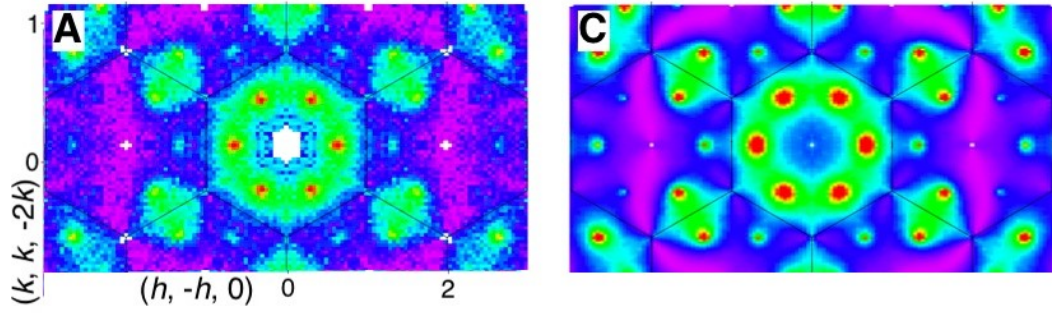

Figure 8. (A) Neutron measurements of the diffuse magnetic scattering in the kagomé spin-ice compound $Dy_2Ti_2O_7$ at T=0.43 K and B=0.5 T. The sharp structural Bragg peaks, such as (2,-2,0), are contained within one pixel and have been removed from the plot. (C) Monte Carlo simulations of the expected scattering in this kagomé spin-ice state. The overall features are in excellent agreement with the data (adapted from [41], © The Physical Society of Japan 2009).

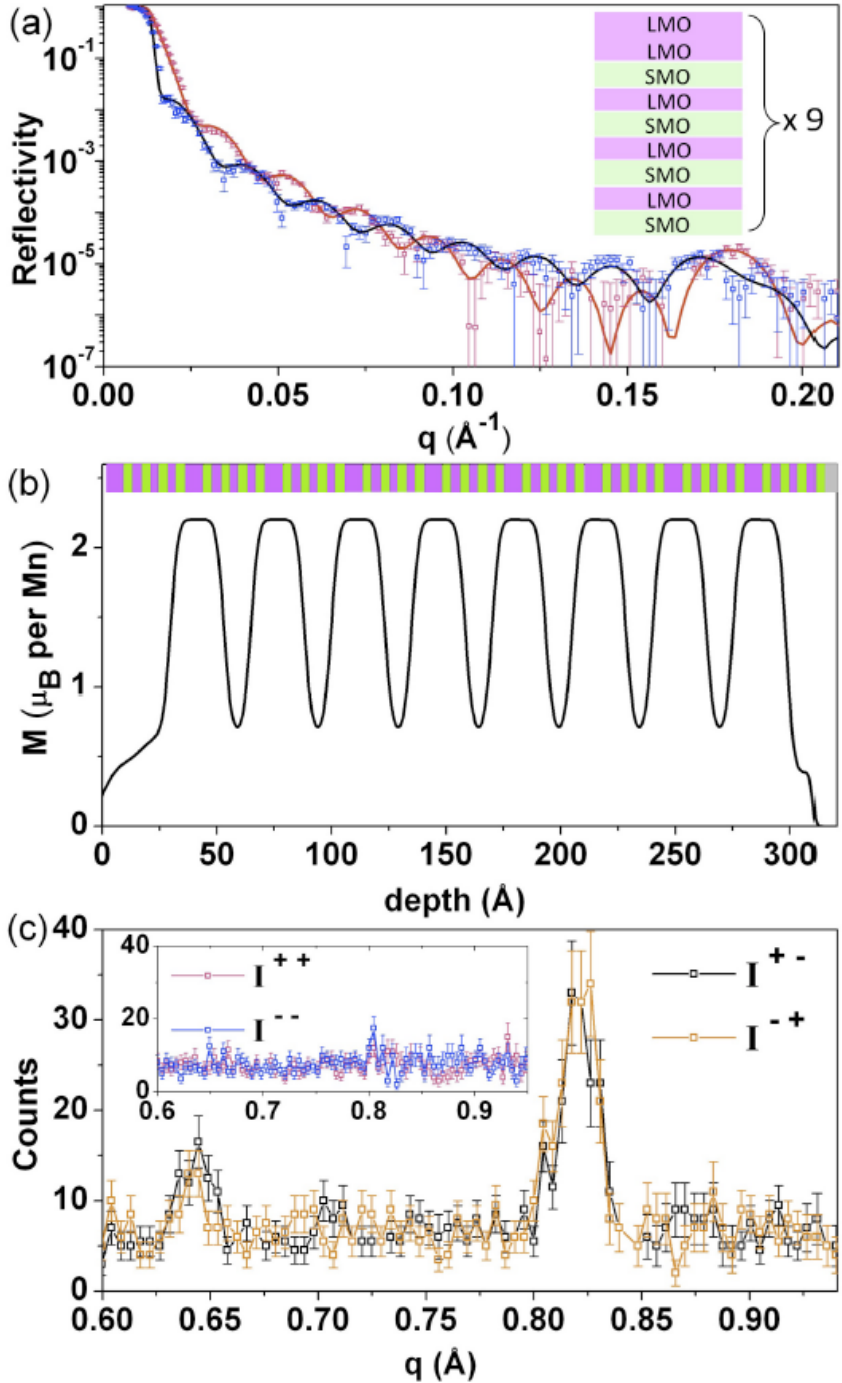

Figure 9. (a) Non-spin-flip polarized neutron reflectivity data $R^{++}$ (red) and $R^{--}$ (blue) on a LaMnO$_3$/SrMnO$_3$ multilayer, measured in a 675 mT field at 120 K. The inset shows a schematic of the superlattice. (b) Magnetic depth profile determined by the fit to the data. Location of the LaMnO$_3$ (pink) and SrMnO$_3$ (green) layers are shown. (c) Spin-flip intensity, showing the antiferromagnetic peak and satellite peak. Inset shows the non-spin-flip scattering in the same range (adapted from [45], © American Physical Society 2011).

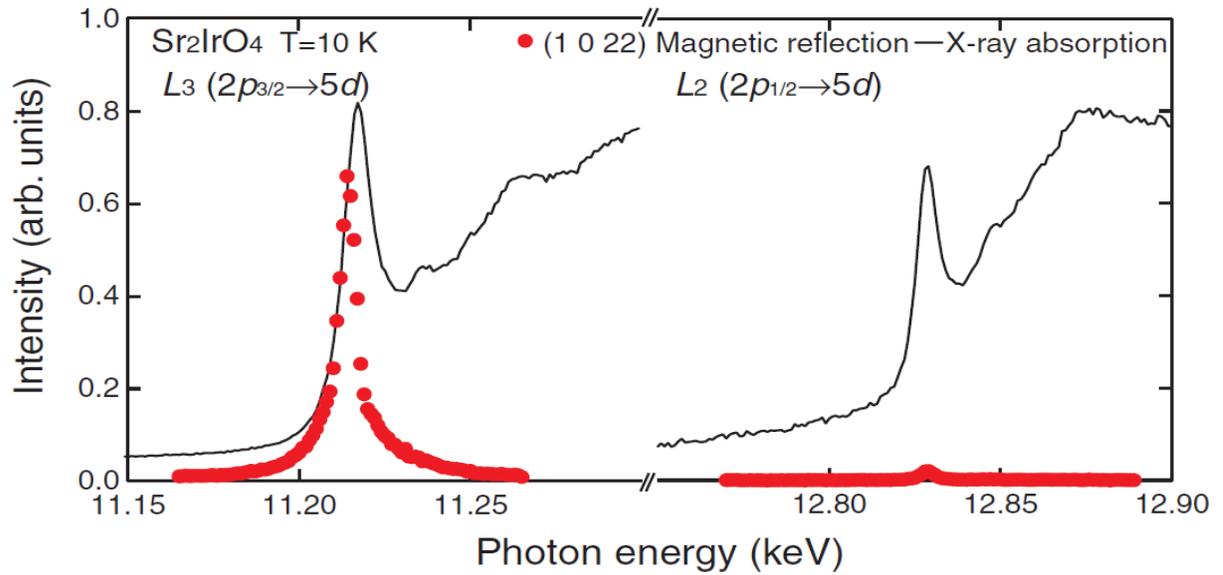

Figure 10. Photon energy dependence of the (1, 0, 22) magnetic Bragg reflection at the $L_3$-(left) and $L_2$-edges (right) of $Sr_2IrO_4$. The black lines show the x-ray absorption spectra for comparison (reprinted with permission from [46], © American Association for the Advancement of Science 2009).

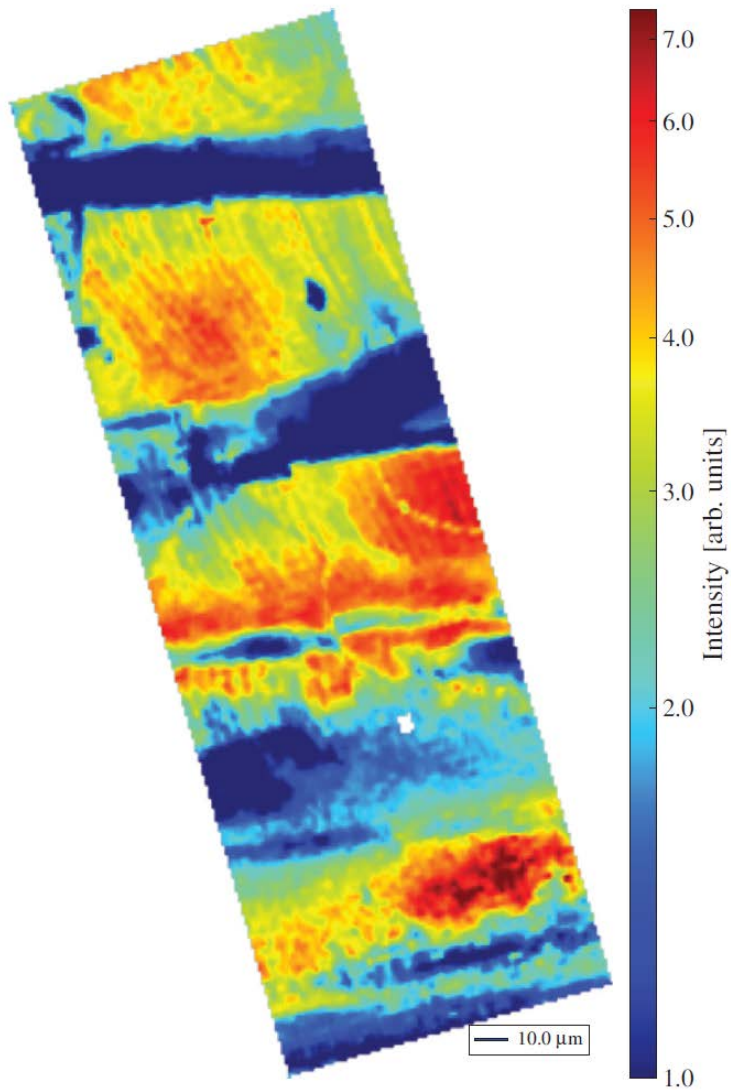

Figure 11. Map of the resonant elastic x-ray scattering intensity at the (0, 0, 1) magnetic Bragg reflection of $La_{0.96}Sr_{2.04}Mn_2O_7$ at the Mn $L_3$-edge. The data indicate domains where the Mn spins point in different directions in the $MnO_2$ layers (reprinted with permission from [47], © American Physical Society 2013).

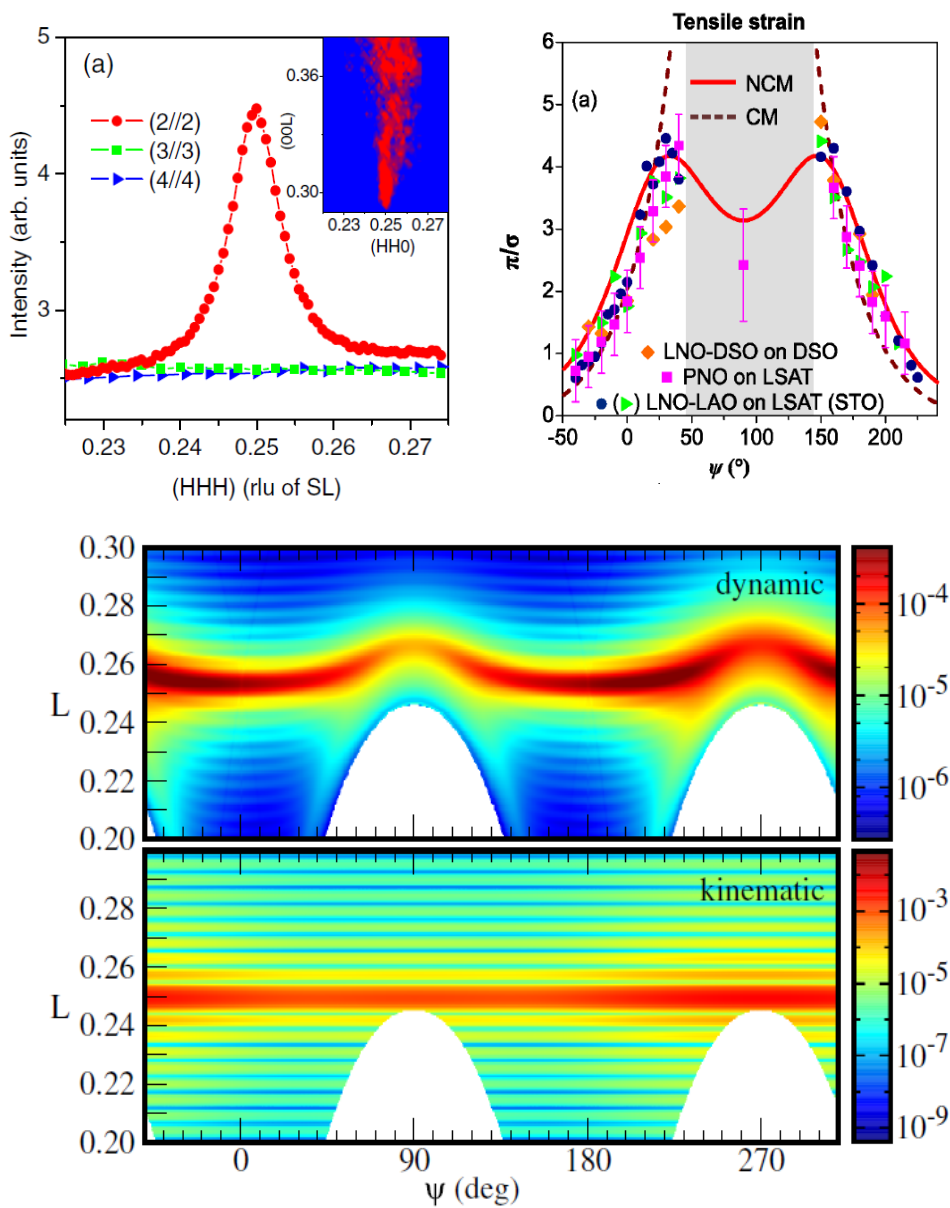

Figure 12. (top left) Ni $L_3$-edge scans through the (¼, ¼, ¼) magnetic Bragg reflection of LaNiO$_3$-LaAlO$_3$ superlattices with different numbers of consecutive unit cells. The absence of the magnetic Bragg peak in superlattices with 3 or more LaNiO$_3$ layers indicates that the magnetic order in the 2x2 superlattice is induced by spatial confinement of the conduction electrons. (reprinted with permission from [48]) (top right) Azimuthal angle dependence of the (¼, ¼, ¼) magnetic Bragg peak of nickelate thin films and superlattices with simulations that rule out collinear (CM) and favor non-collinear (NCM) magnetism. (bottom) Simulated contour map of the scattering intensity of 2x2 LaNiO$_3$-LaAlO$_3$ superlattice as functions of azimuthal angle and momentum transfer perpendicular the to superlattice plane, demonstrating the importance of dynamical diffraction effects (reprinted with permission from [49], © American Physical Society 2016).

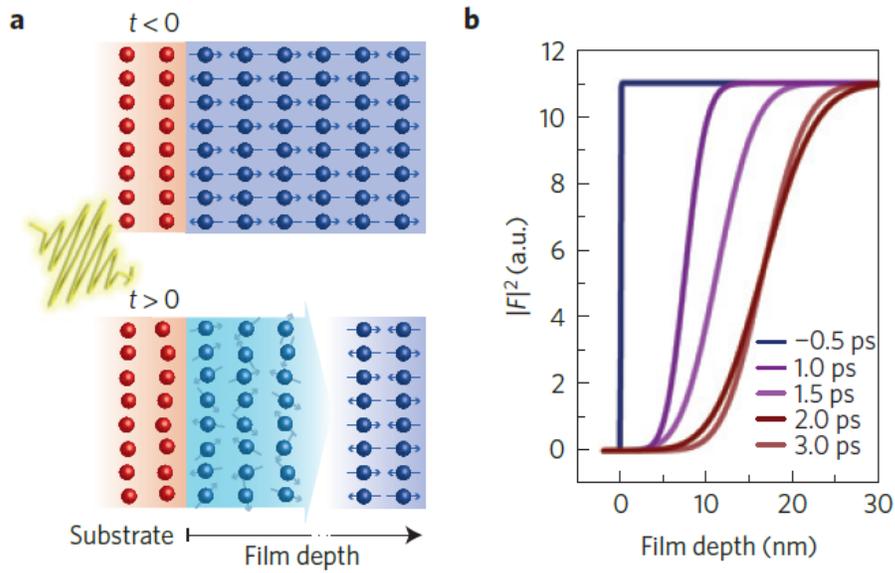

Figure 13. (a) Schematic illustration of the demagnetization process of a NdNiO$_3$ film triggered by a coherently excited photon of the LaAlO$_3$ substrate. (b) Depth profile of the (¼, ¼, ¼) resonant magnetic Bragg peak intensity at different time delays between the phonon pump pulse and the resonant x-ray diffraction probe measurement. (reprinted with permission from [50], ©Springer Nature 2015).

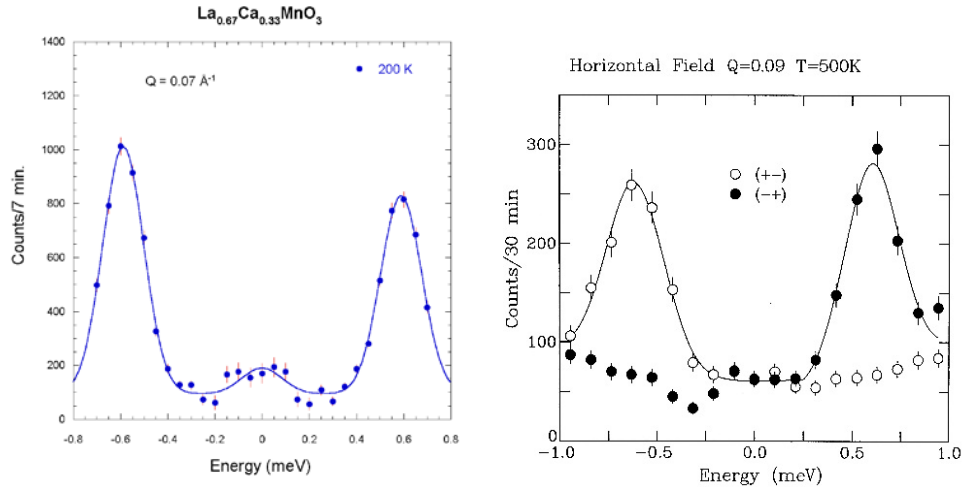

Figure 14. Spin waves in isotropic ferromagnets. (left) Energy scan at a wave vector **q** of 0.07 Å$^{-1}$ for $La_{0.7}Ca_{0.3}MnO_3$, (published with permission from [51], © American Physical Society 1996) showing the spin waves in energy gain (E<0) and energy loss (E>0). (Right) polarized beam energy scan on the $Fe_{86}B_{14}$ amorphous ferromagnet at a fixed wave vector of 0.09 Å$^{-1}$, with the neutron polarization parallel to **q**. In this configuration spin angular momentum is conserved, and the neutron can only create an excitation (E>0) if its moment is initially antiparallel to the magnetization, and can only destroy a spin wave (E<0) when its moment is parallel (reprinted with permission from [52], © American Institute of Physics 1996).

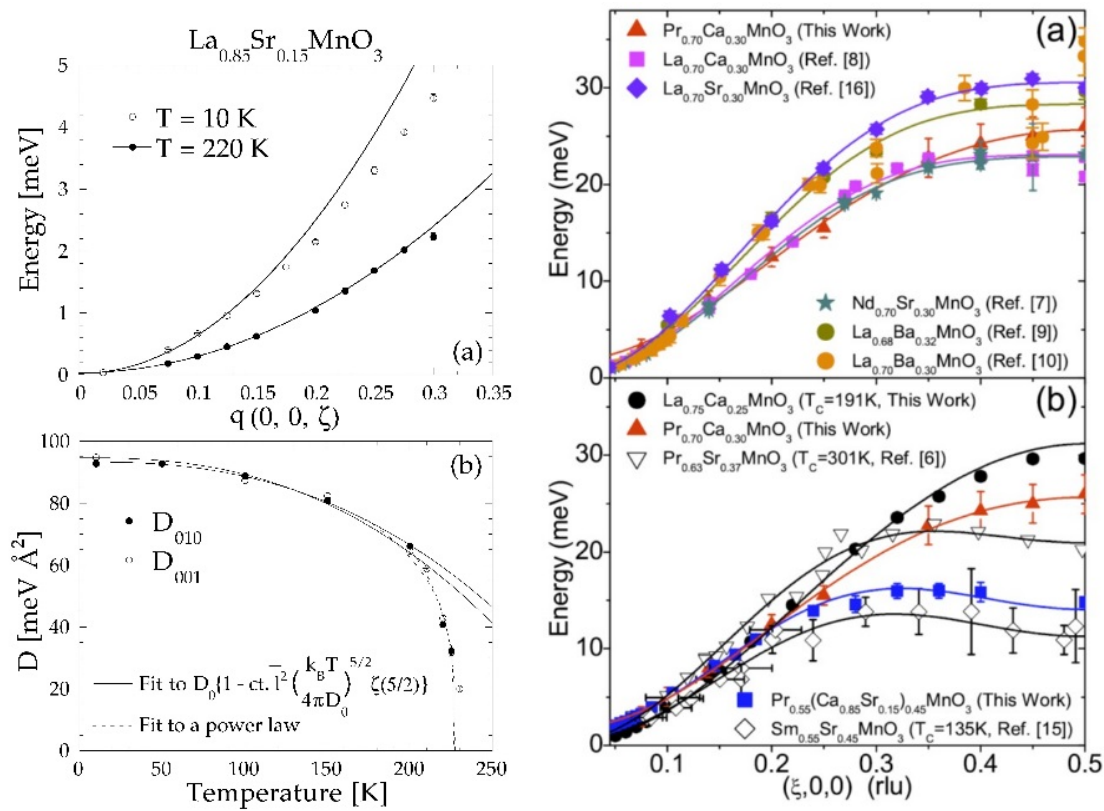

Figure 15. (a) Low energy spin wave dispersion relations at two different temperatures. The dispersion relation follows a quadratic dependence expected for a ferromagnet, which defines the spin stiffness $D$, and no significant gap in the excitation spectrum is observed indicating an isotropic system. $D(T)$ is shown in (b), which follows a power law behavior as the Curie temperature is approached (reprinted with permission from [53], © American Physical Society 1998). (right) Spin wave dispersion relations for a series of colossal magnetoresistive perovskite oxides (reprinted with permission from [54], © American Physical Society 2006).

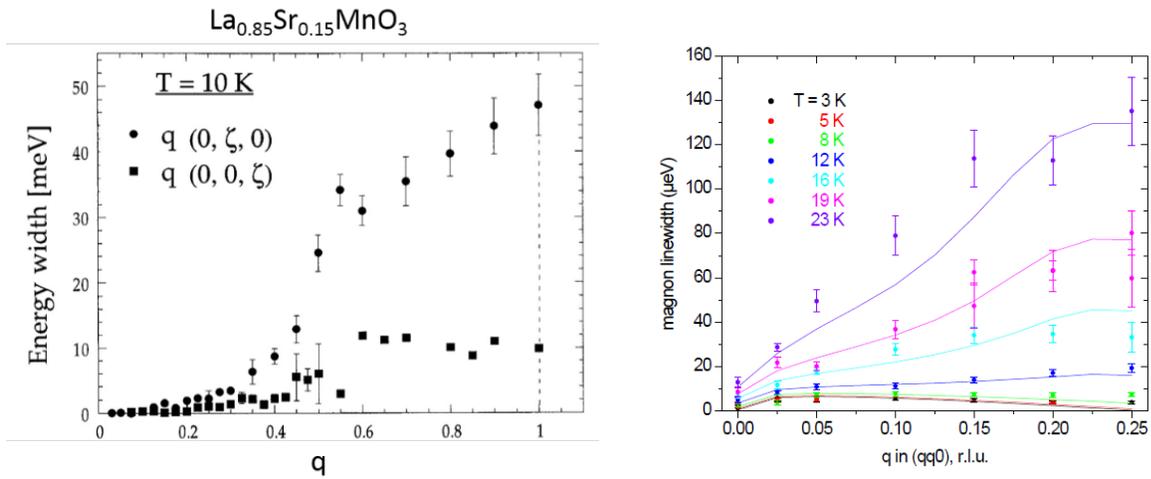

Figure 16. (left) Intrinsic spin wave linewidths for the ground state magnetic excitations in La$_{0.85}$Sr$_{0.15}$MnO$_3$. The linewidths are quite anisotropic, and are significant at small wave vectors but become very large at large *q* (reprinted with permission from [53], © American Physical Society 1998). (Right) magnon linewidths as a function of temperature for a series of *q*'s in the insulating antiferromagnet Rb$_2$MnF$_4$, measured using the high resolution spin-echo triple-axis technique (reprinted with permission from [55], © American Physical Society 2006). The solid curves are calculations using spin wave theory.

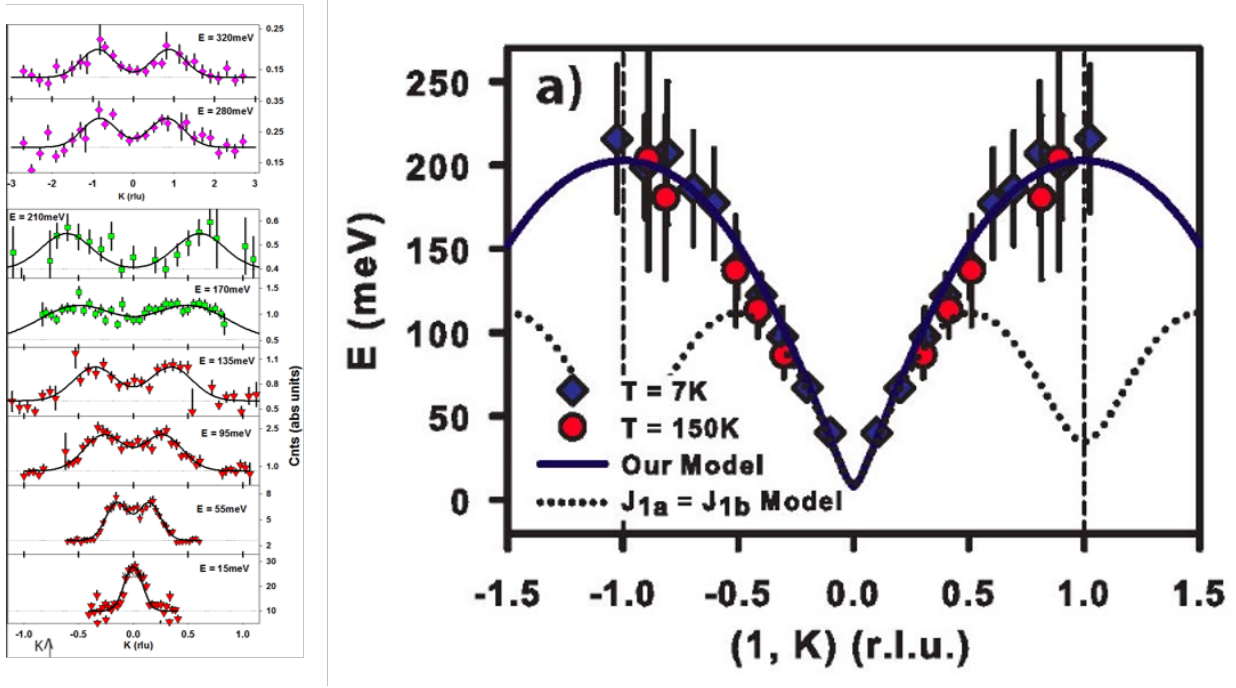

Fig. 17. Left: Constant-energy cuts of the magnetic excitations in BaFe$_2$As$_2$ at a series of energies. The solid curves are the fits to the spin wave model. (Right) Spin wave dispersion along the (1, $K$) direction as determined by energy and $Q$ cuts of the raw data. The solid line is a Heisenberg model calculation using anisotropic exchange couplings $SJ1a = 59.2 \pm 2.0$, $SJ1b = -9.2 \pm 1.2$, $SJ2 = 13.6 \pm 1.0$, $SJc = 1.8 \pm 0.3$ meV determined by fitting the full cross section. The dotted line is a Heisenberg model calculation assuming isotropic exchange coupling $SJ1a = SJ1b = 18.3 \pm 1.4$, $SJ2 = 28.7 \pm 0.5$, and $SJc = 1.8$ meV (adapted from [58], © American Physical Society 2011).

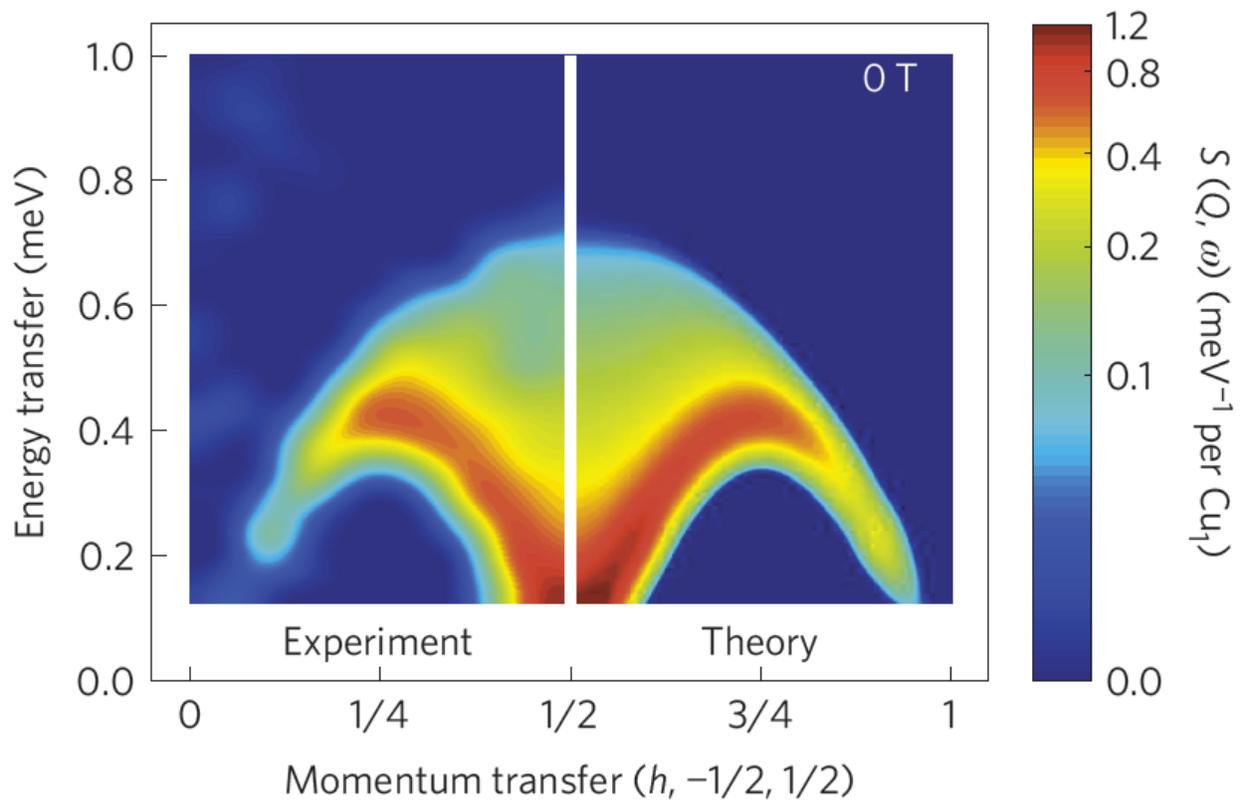

Fig. 18. Intensity color maps of the experimental inelastic neutron scattering spectrum measured along the Cu chain in $CuSO_4 \cdot 5D_2O$ are shown in the left, compared with the theoretical two- and four-spinon dynamic structure factor (reprinted with permission from [60], © Springer Nature 2013).

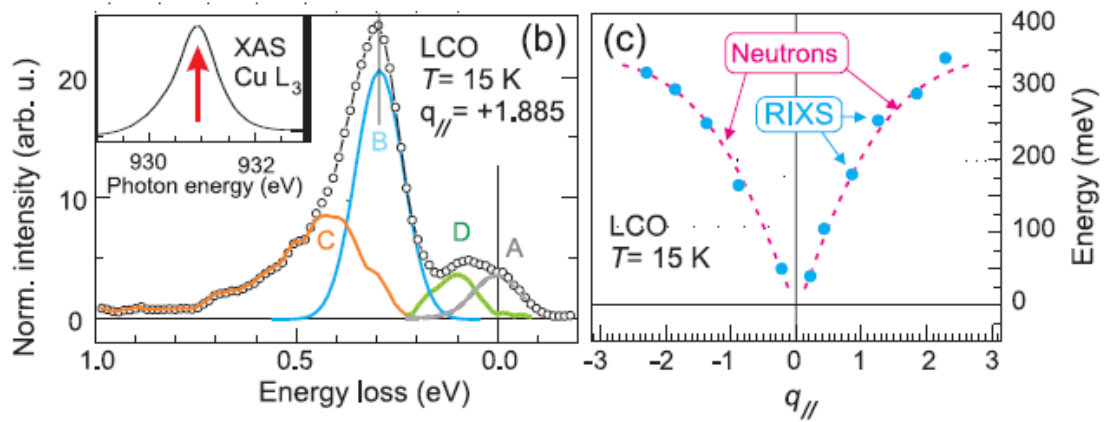

Figure 19. (left) RIXS profile of $La_2CuO_4$ taken at the Cu $L_3$ edge with ≈100 meV energy resolution. The spectrum can be decomposed into elastic (A), single magnon (B), multiple magnon (C) and optical phonon (D) components. The inset shows the x-ray absorption spectrum near the Cu $L_3$ edge, the arrow marks the energy of the incident photons. (right) Single magnon dispersion determined by RIXS (blue dots), compared to inelastic neutron scattering data (dashed line) (reprinted with permission from [23], © American Physical Society 2010).

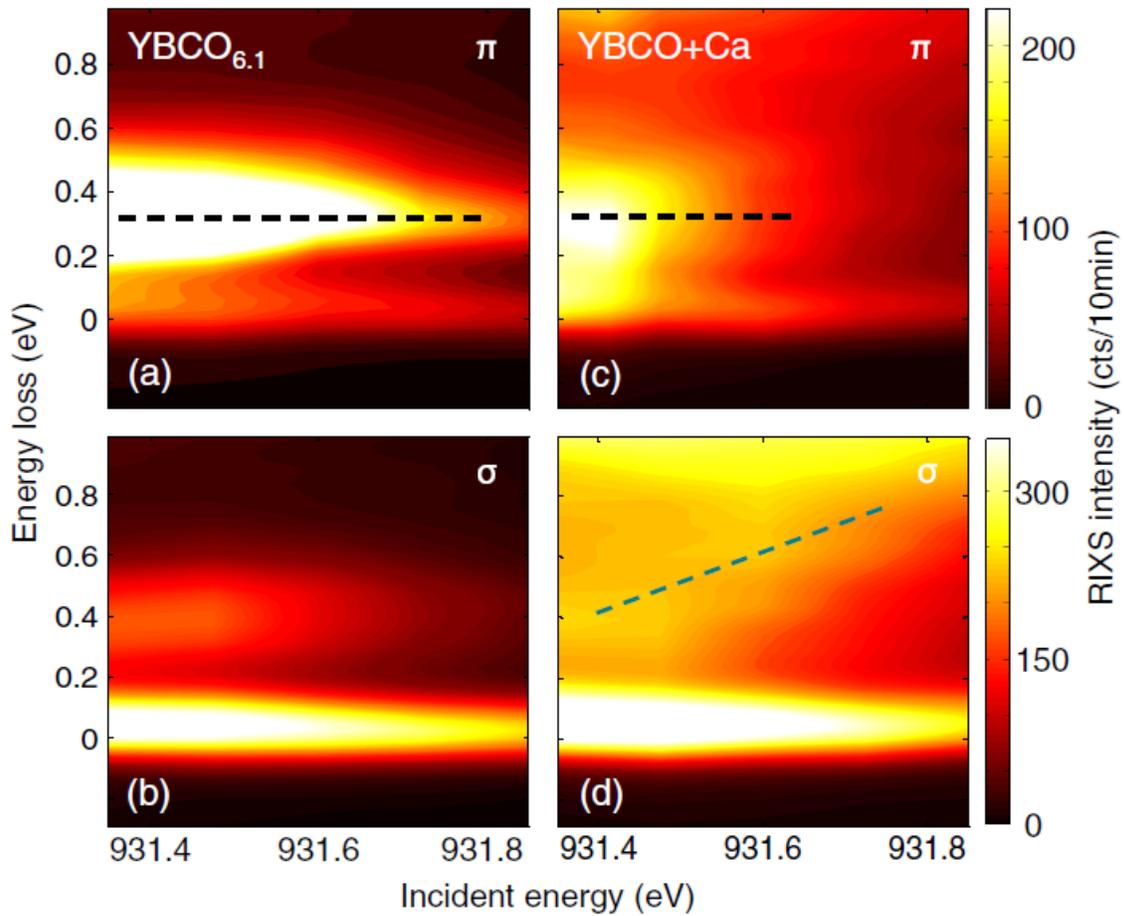

Figure 20. Photon energy dependence of the RIXS intensity ((a),(b)) for undoped antiferromagnetic YBa$_2$Cu$_3$O$_{6.1}$ and ((c), (d)) superconducting Ca-substituted YBa$_2$Cu$_3$O$_7$ in polarization geometries that predominantly select spin (a),(c) and charge (b),(d) excitations. The horizontal dashed lines highlight the energy independence of the magnetic peak position, while the dashed green line is a guide to the eye underlining the fluorescence behavior of the continuum of charge excitations from the doped holes (reprinted with permission from [29], © American Physical Society 2015).

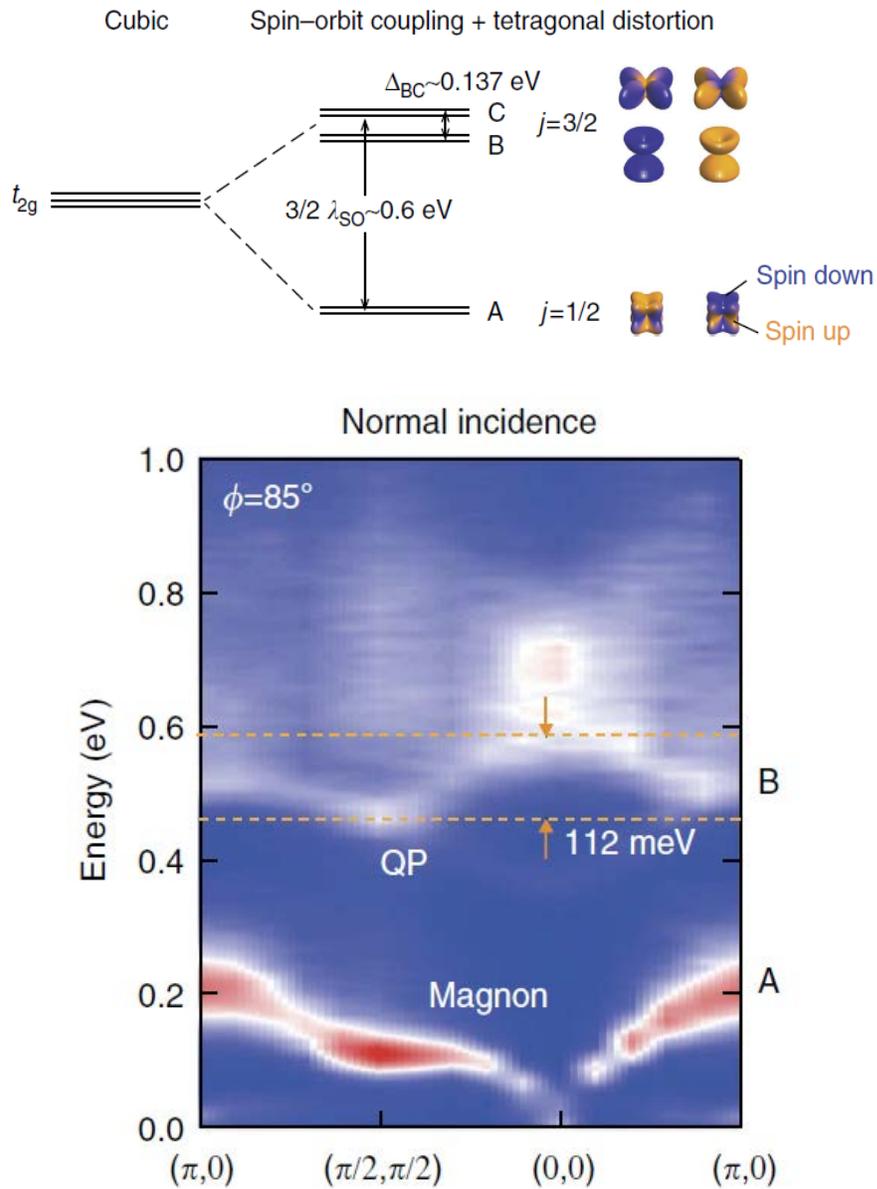

Figure 21. (top) The spin–orbital level scheme of $Sr_2IrO_4$. The spin-orbit coupling $\lambda$ splits the d-electron manifold into $J_{eff}=1/2$ and $3/2$ multiplets. The crystal field $\Delta$ lifts the degeneracy of the $J_{eff}=3/2$ multiplet. Orange (blue) colors in the images of the orbitals represent spin up (down) projections. (bottom) Dispersion of magnons and spin-orbit excitons (marked with QP for "quasiparticle") extracted from RIXS data at the Ir L-edge (reprinted with permission from [27], © Springer Nature 2014).

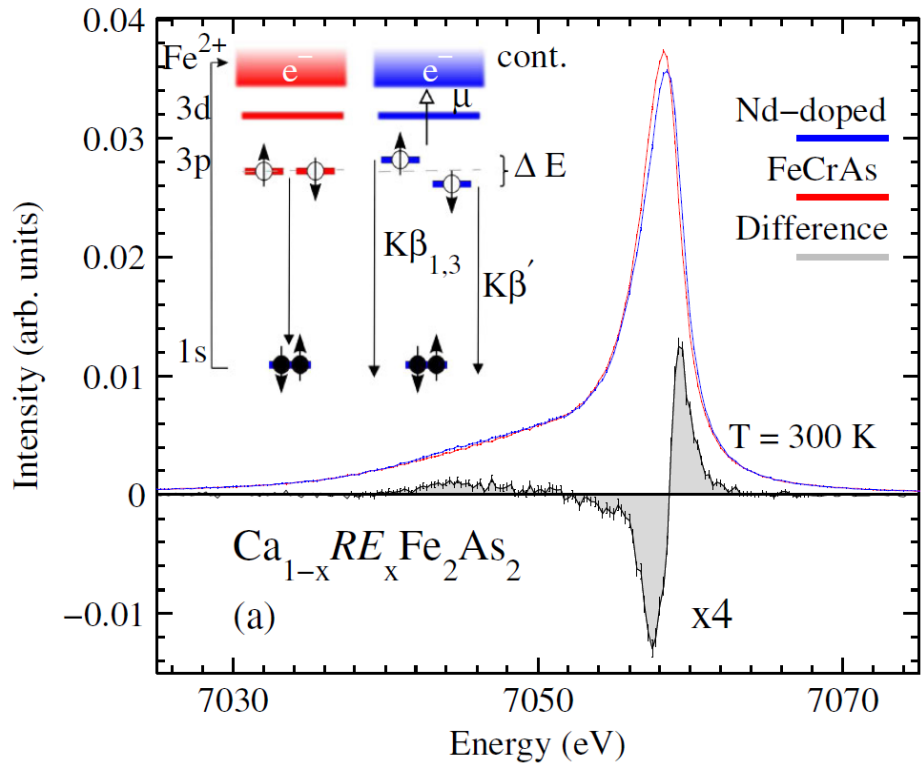

Figure 22. Fe K$_\beta$ emission spectra of Ca$_{1-x}$R$_x$Fe$_2$As$_2$ with $R$ = Nd, and difference spectrum with FeCrAs where Fe is in a nonmagnetic spin-0 state. The difference spectrum indicates a splitting of the emission line due to a local magnetic moment on the Fe site (inset) (reprinted with permission from [61], © American Physical Society 2013).